\documentclass[12pt]{iopart}
\usepackage{graphicx}
\usepackage{iopams}
\usepackage{txfonts}
\usepackage{hyperref}

\newcommand{\sgn}{\textrm{sgn}}                
\def\bra#1{\mathinner{\langle{#1}|}}           
\def\ket#1{\mathinner{|{#1}\rangle}}           
\def\braket#1{\mathinner{\langle{#1}\rangle}}  

\def\numpartsapp{\addtocounter{equation}{1}%
     \setcounter{eqnval}{\value{equation}}%
     \setcounter{equation}{0}%
     \def\theequation{\ifnumbysec
     \Alph{section}.\arabic{eqnval}{\it\alph{equation}}%
     \else\arabic{eqnval}{\it\alph{equation}}\fi}}

\def\endnumpartsapp{\def\theequation{\ifnumbysec
     \Alph{section}.\arabic{equation}\else
     \arabic{equation}\fi}%
     \setcounter{equation}{\value{eqnval}}}

\eqnobysec

\begin{document}

\title[Adapted CUT to treat systems with
  quasiparticles of finite lifetime]{Adapted continuous unitary transformation to treat systems with
  quasiparticles of finite lifetime}

\author{Tim Fischer, Sebastian Duffe and G\"otz S Uhrig}

\address{Lehrstuhl f\"ur theoretische Physik I, Otto-Hahn-Stra\ss e 4,
  D-44221 Dortmund, Germany}

\eads{\mailto{fischer@fkt.physik.tu-dortmund.de},
  \mailto{duffe@fkt.physik.tu-dortmund.de}, \mailto{goetz.uhrig@tu-dortmund.de}}

\begin{abstract}
  {An improved generator for continuous unitary transformations is introduced
    to describe systems with unstable quasiparticles. Its general properties
    are derived and discussed. To illustrate
    this approach we investigate the asymmetric antiferromagnetic
    spin-${1}/{2}$ Heisenberg ladder, which allows for
    spontaneous triplon decay. We present results for the low
    energy spectrum and the momentum resolved spectral density of this
    system. In particular, we show the resonance behavior of the decaying
    triplon explicitly.}
\end{abstract}




\maketitle

\section{Introduction}
\label{sec:introduction}

Most low-energy properties of condensed matter systems can be understood in
terms of quasiparticles, which are the elementary excitations of a
system. Higher lying excitations are described in terms of scattering states
or bound states of the elementary quasiparticles.
One of the most successful concepts of quasiparticles was developed by Landau 
in the late
1950's for interacting fermionic systems, the so-called 
Fermi liquid theory \cite{Landau_Lifshitz_9-2}. It is
based on the idea that complex interacting excitations are adiabatically
linked to the non-interacting ones. They share the same quantum numbers,
i.e. momentum and spin.

\begin{figure}[ht]
  \centering
  \includegraphics[width=0.65\columnwidth]{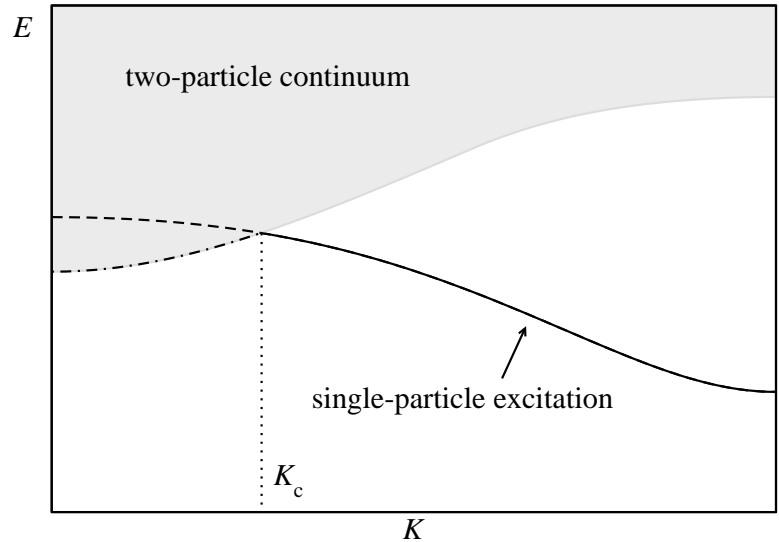}
  \caption{States with two excitations lying energetically below the single 
    particle dispersion for momentum $K<K_{\textrm{c}}$. If in addition the 
    Hamiltonian contains matrix
    elements which connect the one-particle space with the two-particle space,
    the (quasi)particles will become unstable for $K\leq
    K_{\textrm{c}}$ (dashed line). The generator $F_{\textrm{pc}}(l)$ leads to
    the dispersion consisting of the solid and the dashed-dotted lines
    (cf.\ \sref{subsec:eta_pc}).}
  \label{fig:decay}
\end{figure}

Only a few years later, Pitaevskii predicted that these quasiparticles can
become unstable if certain decay channels exist \cite{P59}. 
The quasiparticles do not
survive beyond a certain threshold in momentum space and their spectrum 
terminates at this threshold (see \fref{fig:decay}).
This prediction was confirmed later by neutron scattering
measurements on superfluid $^4$He \cite{SCW77,FB98}. 

In 2006 such a quasiparticle breakdown was measured for the first time
in a quantum magnet \cite{Stone06,MZMRCQ06}. The spin excitations in the 
two-dimensional spin-${1}/{2}$ quantum magnet
piperazinium hexachlorodicuprate (PHCC) show remarkable similarities with the
excitations in superfluid $^4$He. Stone \textsl{et al.} observed a threshold 
momentum beyond which the quasiparticle merges with the two-quasiparticle 
continuum and ceases to exist as well-defined excitation \cite{Stone06}. 
This phenomenon was also observed in the quasi-one-dimensional antiferromagnet 
IPA-CuCl$_3$ by Masuda \textsl{et al.} \cite{MZMRCQ06}.
In theoretical considerations, magnons decaying into pairs of magnons
are found in the long-range ordered  Heisenberg model on the
triangular lattice \cite{zheng06a,zheng06b,chern09} as well as in the
Heisenberg model on square lattices in strong magnetic fields \cite{ZC99}.

Generally, physical systems of unstable quasiparticles are much more common 
than systems of completely stable quasiparticles. For instance, in the case of 
Fermi liquids the
quasiparticles always have a finite lifetime except at the Fermi energy.
For the theoretical description and the understanding of many systems
in condensed matter physics it is therefore an indispensable task to develop 
methods which are able to describe systems with unstable quasiparticles. 

A perturbative analysis of quasiparticle breakdown in quantum magnets was
given in 2006 by Kolezhuk and Sachdev \cite{KS06} and by
Zhitomirsky \cite{Zhi06} based on fully diagrammatic approaches.
Both papers show that elementary excitations in gapped spin systems 
become unstable if they merge with the two-particle continuum.
Spin systems in one and higher dimension are analyzed to explain the
observations in IPA-CuCl$_3$ and  in PHCC.
In one dimension a square-root
dependence of the inverse quasiparticle lifetime is predicted \cite{Zhi06}.
For the special case of an asymmetric rung-dimerized spin ladder
Bibikov \cite{Bibikov07} confirmed these results by Bethe ansatz.
An alternative approach to derive the lifetime of an excitation based
on renormalization group methods was developed by Bach \textsl{et al.}
\cite{BFS98}.

Here we introduce an advancement of the method of continuous
unitary transformations (CUTs) introduced in 1994  by
Wegner \cite{weg94} and independently by G{\l}azek and
Wilson \cite{GW93,GW94}, which allows us to describe systems with quasiparticle
decay. 

The paper is divided into two main parts. In the first part in \sref{sec:method}, we give a short introduction to the method of CUTs
and describe generally how one can
deal with quasiparticle decay within this framework.
In the second part consisting of \sref{sec:model} and
\sref{sec:results}, we illustrate the general concept by
 explicit results for the asymmetric antiferromagnetic 
spin-${1}/{2}$ Heisenberg ladder.

\section{Method: Continuous unitary transformations}
\label{sec:method}
In this section, we first outline the general concept of CUTs. Then we discuss 
similarities and differences between
various schemes of CUTs depending on the choice of the generator. 

In principle, any Hamiltonian can be diagonalized by a suitable unitary
transformation $U$. Famous examples are bosonization \cite{DS98_2} or 
Bogoliubov transformations \cite{BR86} whose
fermionic version is used in the BCS theory of superconductivity \cite{BCS57}.
For complex problems it is usually a very hard task to find such a
suitable transformation. In 1994 Wegner \cite{weg94} (and independently 
G{\l}azek and Wilson \cite{GW93,GW94}) presented a method to diagonalize a 
given Hamiltonian $H$ in a continuous way. This method of CUTs is based on the 
idea to introduce a continuous auxiliary variable $l$ and to define an 
$l$-dependent Hamiltonian $H(l):=U^{\dag}(l)H U(l)$. Then the Hamiltonian 
transforms  according to the flow equation
\begin{equation}
  \label{eq:flow_equation}
  \partial_l H(l) = \left[ F(l), H(l) \right],
\end{equation}
with an anti-Hermitian generator $F(l):= -U^{\dag}(l)
\left( \partial_l U(l) \right)$. For $l \rightarrow \infty$ 
the flow equation \eref{eq:flow_equation} maps the initial Hamiltonian 
$H(0):=H$ to an effective Hamiltonian $H_{\textrm{eff}}:=H(\infty)$ in a
unitary way. Certainly, the final structure of the effective Hamiltonian
$H_{\textrm{eff}}$ depends on the form of the chosen generator $F(l)$. So the 
crucial point is to choose a generator $F(l)$ which leads to a 
simplification of the initial Hamiltonian. Another important issue is whether 
the ensuing flow
equation \eref{eq:flow_equation} is practically tractable.

Wegner proposed to define the generator $F(l)$ as the commutator between
the diagonal part of the Hamiltonian $H_\textrm{d}(l)$ and the Hamiltonian
$H(l)$ itself. So the generator reads $F(l) = 
\left[H_{\textrm{d}}(l), H(l) \right]$.
It was proven~{\cite{weg94,DU04}} that this choice transforms the
Hamiltonian in such a manner that $\left[H_{\textrm{d}}(\infty), H(\infty)
\right]=0$, which implies that the final Hamiltonian $H(\infty)$ is
block-diagonal with respect to the eigensubspaces of
$H_{\textrm{d}}(\infty)$. If $H_{\textrm{d}}(\infty)$ is non-degenerate the
final Hamiltonian $H(\infty)$ is actually diagonal.

For band-diagonal Hamiltonian matrices, Mielke proposed 
another generator. His choice conserves the initial band structure during the
flow~{\cite{Mie98}}, which is not the case for Wegner's
generator. Mielke achieved the conservation of the band structure by
introducing a sign function depending on the difference between the row index
and the column index of the considered matrix element.

Independently thereof, Knetter and Uhrig~{\cite{UN98,KU00}} suggested a 
generator which allows us to
create (quasi)particle number conserving effective many-body
Hamiltonians. Their choice is
also based on the idea to use a sign function. In contrast to Mielke's
choice they used the difference of the particle number as the argument of the
sign function. This generator can be regarded as a generalization of Mielke's 
generator for Hamiltonians formulated in second quantization.
In the following, we denote this generator creating (quasi)particle number
conserving effective Hamiltonians by $F_{\textrm{pc}}(l)$.
An analogous generator was also used by Stein~{\cite{S97,S98}}
for models where the use of the sign function was not necessary.

In the following, we first summarize some properties of the
generator $F_{\textrm{pc}}(l)$ and specify its pros and cons. 
Particularly, we describe the
problems arising in the description of systems with unstable (quasi)particles.
Thereafter we present possible variations of the generator
$F_{\textrm{pc}}(l)$ including a generator which allows for the description 
of (quasi)particles with finite lifetime.

\subsection{The generator $F_{\textrm{pc}}(l)$}
\label{subsec:eta_pc}
Generally, a Hamiltonian in second quantization can be written as
\begin{equation}
  \label{eq:hamiltonian}
  H(l) = \sum_{i,j=0}^{N} H_{j}^{i}(l),
\end{equation}
where $H_{j}^{i}(l)$ stands for the sum over all normal ordered terms which 
create $i$ and annihilate $j$ (quasi)particles, e.g. $H^0_0(l)$ is proportional
 to the identity and describes the vacuum energy during the flow.
By the expression ``term'', we refer to both, the operators and
the corresponding prefactor. The whole $l$-dependence of the Hamiltonian is 
carried by the prefactors.
Note that for infinitely large systems the maximum
number of involved quasiparticles $N$ may be infinite, but 
this does not need to be the case. 

According to the form of the Hamiltonian \eref{eq:hamiltonian} the generator 
$F_{\textrm{pc}}(l)$ is given by 
\begin{equation}
  \label{eq:eta_pc}
  F_{\textrm{pc}}(l) = \sum_{i,j=0}^{N} \sgn \left(i - j \right) H_{j}^{i}(l).
\end{equation}
This means that terms in $H(l)$ which contain more creation operators than
annihilation operators are taken over to $F_{\textrm{pc}}(l)$ with the same 
sign. Terms with more annihilation operators than creation operators are 
included in $F_{\textrm{pc}}(l)$ with a negative sign. Terms leaving the 
number of particles unchanged do not occur in $F_{\textrm{pc}}(l)$.
 
For the generator $F_{\textrm{pc}}(l)$ the flow equation 
\eref{eq:flow_equation} exhibits the following properties:
\begin{itemize}
  \item[a)] If the spectrum of $H$ is bounded from below, the flow equation
    converges \cite{Mie98,KU00}. This is the generic situation for physical
    systems. The mathematical derivation requires the Hilbert space of the
    system to be finite dimensional.
  \item[b)] The effective Hamiltonian $H_{\textrm{eff}}$ is block-diagonal in
    the sense that it conserves the (quasi)particle number \cite{KU00}. 
    Therefore, the effective Hamiltonian commutes with the operator $Q$ which
    counts the number of (quasi)particles
    \begin{equation}
      \left[H_{\textrm{eff}},Q\right] = 0.
    \end{equation}
    Thus it is of the form
    \begin{equation}
      H_{\textrm{eff}} = \sum_{i=0}^{N} H_{i}^{i}(\infty).
    \end{equation}
    This property allows us to analyze subspaces with different (quasi)particle
    numbers separately.
  \item[c)] If the initial Hamiltonian $H(0)$ has a block band-diagonal
    structure (i.e., $H_{j}^{i}(0) = 0$ for $\left|i-j\right|>N_0$), this
    block band-diagonal structure is conserved during the
    flow \cite{Mie98,KU00}.
  \item[d)] The generator $F_{\textrm{pc}}(l)$ sorts the eigenvalues in 
    ascending order of the particle
    number of the corresponding eigenvectors \cite{Mie98,Heidbrink2002} if
    the eigenvectors are linked by a matrix element of the Hamiltonian (see
    also \sref{app:ordering}).
\end{itemize} 

\begin{figure*}
  \includegraphics[width=1.0\textwidth]{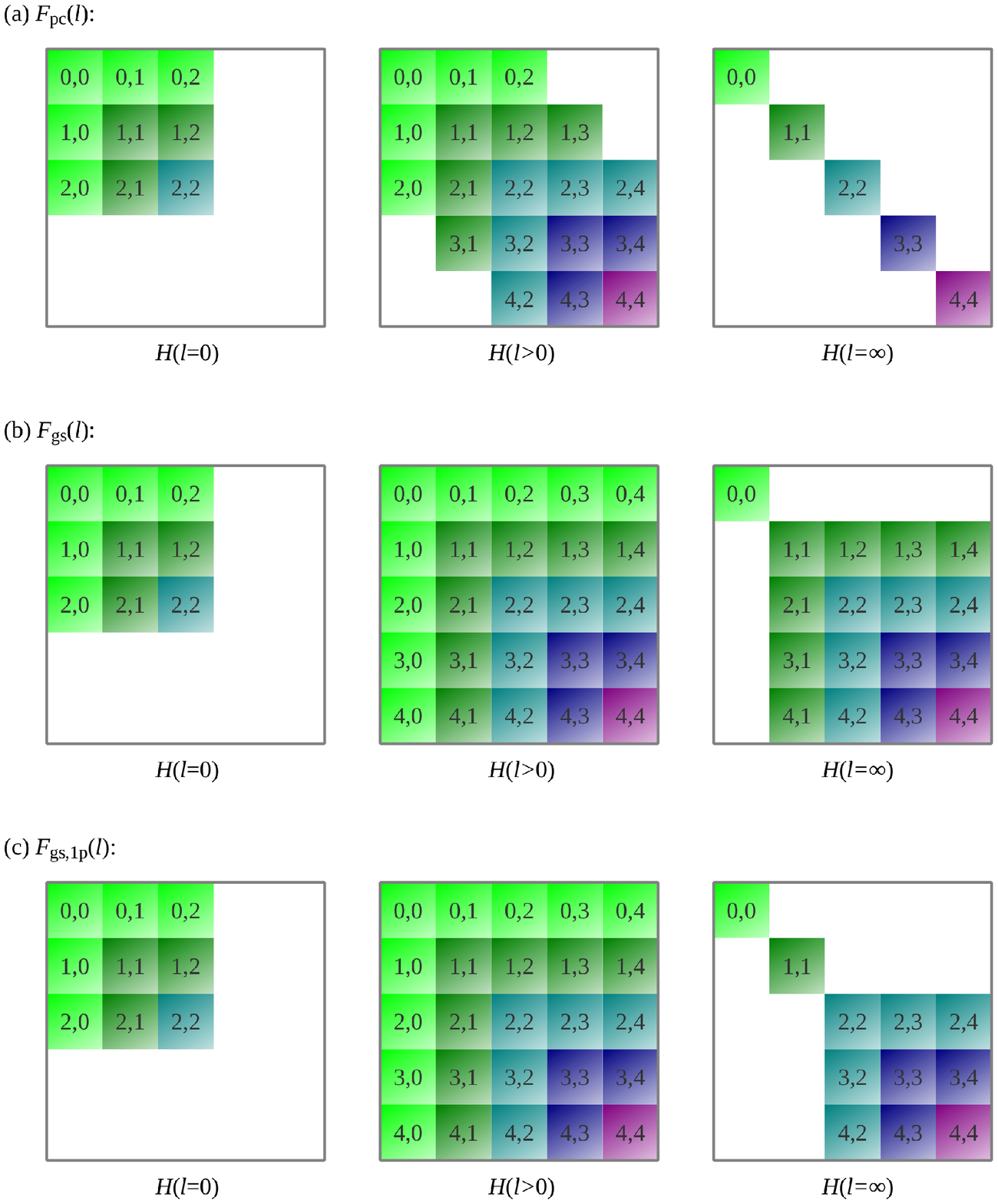}
  \caption{Schematical representation of the structure of the Hamiltonian
      $H(l)$ during the flow for various generators. A colored block
      described by the pair $i,j$ stands for the part $H^i_j(l)$ of the
      Hamiltonian. Only those blocks are colored where at least one term of
      $H^i_j(l)$ has a non-vanishing coefficient. In all cases, we assume an
      initial Hamiltonian which creates or annihilates at most two
      particles. For simplicity, we restricted our illustrations to terms which
      at most create or annihilate four particles for $l>0$. Of course, terms 
      which create or annihilate more than four
      terms may also occur. Panel (a) shows that the generator
      $F_{\textrm{pc}}(l)$ conserves the block band-diagonality of the
      initial Hamiltonian during the flow and leads to a (quasi)particle number
      conserving effective Hamiltonian. Panels (b) and (c) show that both
      generators, $F_{\textrm{gs}}(l)$ and $F_{\textrm{gs,1p}}(l)$, do not
      conserve the block band-diagonality. The generator
      $F_{\textrm{gs}}(l)$ only separates the $H^0_0(l)$ part, whereas the
      generator $F_{\textrm{gs,1p}}(l)$ also decouples the $H^1_1(l)$ part.} 
  \label{fig:comparison}  
\end{figure*}

Items b) and c) are schematically illustrated in \fref{fig:comparison}.
Despite all the favorable properties of the generator $F_{\textrm{pc}}(l)$, it
is not advantageous in every situation. 
Particularly, the last point is both a blessing and a curse. On the one hand,
it ensures that the ground state is represented by the vacuum state of
the effective model\footnote{For simplicity, the ground state
  $\ket{0}$ is assumed to be unique.}. Additionally, it produces the 
appropriate (quasi)particle
picture in systems where the elementary excitations have an infinite
lifetime. But on the other hand, the described ordering of the eigenstates 
does not reflect the situation in many physical systems, e.g. systems with 
unstable (quasi)particles. This is schematically illustrated in \fref{fig:decay}.

The generator $F_{\textrm{pc}}(l)$ interprets the energetically lowest states 
above the ground state as the elementary excitations. In principle, it is 
possible to define the elementary excitations of the system in this way. 
But this definition can be misleading in the sense that states with
very low or zero spectral weight are regarded as the elementary
excitations of the system. Without spectral weight we consider such
states to be meaningless in terms of elementary excitations which serve
as building blocks of all other excitations. Therefore, one usually defines 
the states with the largest spectral weight above the ground state as the
elementary excitations of the system.
Moreover, previous calculations \cite{RMHU04,Reischl2006} strongly suggest 
that the rearrangement of the
Hilbert space causes convergence problems in practice. In the perturbative
approach of CUT \cite{KU00,KSU03} (p-CUT) these problems become perceivable in
the extrapolations \cite{SU05,DU09}.

The second property, $ \left[H_{\textrm{eff}},Q\right] = 0$, of the effective
Hamiltonian generated by $F_{\textrm{pc}}(l)$  makes the describtion of 
unstable (quasi)particles difficult. By construction, the generator 
$F_{\textrm{pc}}(l)$
produces an effective Hamiltonian where the elementary excitations exhibit an
infinite lifetime. The information of the decay is stored
in the unitary transformation and therefore an additional transformation of
observables is indispensable to describe the quasiparticle decay. 
This approach was first used by Kehrein and Mielke to describe dissipative
quantum systems \cite{KM97,KM98}. 

In the following subsection, we present a generator which does not eliminate the
 decay processes. Therefore, it is possible to study the quasiparticle decay 
more easily and more directly. The transformation of the observable
is still necessary for quantitative results, but the essential aspect, i.e.
the finite life time, is obvious without this transformation.

\subsection{Generator for the ground state}
\label{subsec:gs}
To tackle the problems of
(quasi)particle decay within the framework of CUTs mentioned in the previous 
section we introduce the adapted generator
\begin{equation}
  \label{eq:eta_gs}
  F_{\textrm{gs}}(l)=\sum_{i>0}^N \left(H_{0}^{i}(l) - H_{i}^{0}(l)\right)
\end{equation}
relying on the form of the Hamiltonian \eref{eq:hamiltonian}. We included 
only those terms in the generator $F_{\textrm{gs}}(l)$ which
either contain only creation operators or contain only annihilation
operators. The terms which contain only creation operators are included as they
appear in $H(l)$. The terms which contain only annihilation operators
are included with a negative sign relative to their sign in $H(l)$.

Again, the flow equation \eref{eq:flow_equation} converges if the
spectrum is bounded from below. This follows directly from introducing a basis
$\left\{ \ket{i} \right\}$, including the vacuum state $\ket{0}$, and examining
\begin{equation}
  \label{eq:del_H00}
  \partial_l H_{0,0}(l) = -2 \sum_{i\neq 0} \left|H_{0,i}(l) \right|^2
\end{equation}
with $H_{i,j}(l):=\bra{i} H(l) \ket{j}$. Note that $H_{i,j}(l)$
describes an explicit matrix element in contrast to the previously
appearing quantity $H_{j}^{i}(l)$, which stands for a sum over
terms in second quantization. According to \eref{eq:del_H00},
$ H_{0,0}(l) $ is a monotonically decreasing function of
$l$. Therefore, if the spectrum is bounded from below, its derivative must
vanish in the limit $l \rightarrow \infty$. This also implies that
\begin{equation}
  \lim_{l \rightarrow \infty} H_{0,i} (l) = \lim_{l \rightarrow \infty} 
H_{i,0}^* (l)  = 0,
\end{equation}
i.e., all matrix elements connected to the vacuum state vanish in the limit $l
\rightarrow \infty$. In contrast to the generator $F_{\textrm{pc}}(l)$ this
generator destroys a block band-diagonal structure of the initial
Hamiltonian $H(0)$. It solely separates the vacuum state from all other
states. Hence the effective Hamiltonian is more difficult to analyze. 
This is the consequence of the more
complex physics we have to describe.
The evolution of the Hamiltonian $H(l)$ during the flow
using the generator $F_{\textrm{gs}}(l)$ is compared to the one induced
by $F_{\textrm{pc}}(l)$ in \fref{fig:comparison}.

While the choice \eref{eq:eta_gs} is very plausible, we have not
presented a systematic derivation of
$F_{\textrm{gs}}(l)$ so far.
To provide such an induction we adopt the derivation of a generator in the 
context of variational calculations \cite{DEO08}. The idea of Dawson 
\textsl{et al.} was to minimize $\partial_l E_0(l)= \partial_l \bra{0} H(l)
\ket{0}$
under the constraint of a bounded $F(l)$ so that the quantity $E_0(l)$ 
decreases as fast as 
possible\footnote{To correspond with our approach in second
  quantization we use the vacuum state $\ket{0}$ as the starting
  vector for the minimization. In principle, one can use an arbitrary
  starting vector.}. This leads to the calculation of
\begin{equation}
  \delta \bigg\{ \bra{0} \left[F(l), H(l) \right] \ket{0} +
  \lambda \left\|F(l) \right\|^2_{\textrm{H}} \bigg\}=0
\end{equation}
with the Lagrange multiplier $\lambda >0$ and $\left\|.\right\|_{\textrm{H}}$ 
denoting the Hilbert-Schmidt
norm. With respect to a basis $\left\{ \ket{i} \right\}$, including $\ket{0}$,
one obtains the expression
\begin{equation}
  \delta \Bigg\{ \sum_{i} \bigg(
    F_{0,i}(l)H_{i,0}(l)-H_{0,i}(l)F_{i,0}(l) \bigg)+ \lambda\sum_{i,j}
  \underbrace{F_{i,j}^*(l)}_{-F_{j,i}(l)} F_{i,j}(l) \Bigg\}=0
\end{equation}
with the matrix elements $H_{i,j}(l):=\bra{i} H(l) \ket{j}$ and 
$F_{i,j}(l):=\bra{i} F(l) \ket{j}$. The variation implies
\begin{equation}
   0 = \delta_{0,i} H_{j,0}(l) - H_{0,i}(l)\delta_{j,0} - 2\lambda 
F_{j,i}(l)
\end{equation}
and hence 
\begin{equation}
  F_{i,j}(l) = \frac{1}{2\lambda} \left( H_{i,0}(l) \delta_{0,j} - 
\delta_{i,0}    H_{0,j}(l) \right).
\end{equation}
In the following, we set $\lambda = {1}/{2}$ and denote this
generator by $F_{\textrm{mgs}}(l)$. It has the property that only \emph{matrix
elements} involving the vacuum state $\ket{0}$, i.e. $F_{i,0}(l)$ or
$F_{0,i}(l)$, are different from zero. All other matrix elements vanish.

The appealing property of $F_{\textrm{mgs}}(l)$ is that there is a strong
similarity to $F_{\textrm{gs}}(l)$ in the sense that the terms of
$F_{\textrm{gs}}(l)$ containing only creation
operators (or annihilation operators) represent the matrix elements
$F_{i,0}(l)$ (or $F_{0,i}(l)$) of $F_{\textrm{mgs}}(l)$ among
other processes. But the effect
on the total Hilbert space is very different. The matrix
$F_{\textrm{mgs}}(l)$ is active if and only if there is a direct connection
to the vacuum state $\ket{0}$, while, for instance, a term consisting
only of creation operators also acts on states which already have a
certain number of particles. Therefore, $F_{\textrm{gs}}(l)$ can be seen as a
generalization of $F_{\textrm{mgs}}(l)$ for problems formulated in second
quantization. But $F_{\textrm{gs}}(l)$ and $F_{\textrm{mgs}}(l)$ are
not identical.

The question arises if it is possible to adapt the above variational 
derivation of the generator $F_{\textrm{mgs}}(l)$ to the generator
$F_{\textrm{gs}}(l)$ formulated in second quantization. This can be achieved
by modifying the applied scalar product as we show next.

We consider a system formulated in second quantization.
Each operator acting on the Hilbert space can be
represented by a sum over terms consisting of a product of creation and 
annihilation operators and a prefactor. We call the product of creation and 
annihilation operators
a monomial. Thus a term consists of a monomial and a prefactor.

 To obtain a unique representation of each monomial we first assume them to be
normal ordered. Second, a certain ordering within all creation (annihilation)
operators is implied. The creation and annihilation operators are denoted by 
$e^{\dag}_{i_k}$ and $e^{\phantom{\dag}}_{i_k}$, where $i_k$ contains all 
quantum numbers describing
the considered operator, for instance its position and spin. 
Note that such an expansion of a general operator is unique since all possible 
(ordered)
monomials are linearly independent. They can be distinguished from one another
by appropriate matrix elements. 

Next we define the scalar product of two monomials $M_1$ and $M_2$ by
\begin{equation}	
  \label{eq:scalar-product}
  \left\langle M_1, M_2 \right\rangle :=
  \cases{1 & for $M_1 = M_2$ \\ 0 & for $M_1 \neq M_2$}.
\end{equation}	
Since any operator on the total Hilbert space can be expanded in
monomials, \eref{eq:scalar-product} in combination with the usual
bilinearity of scalar products defines a valid scalar product. The scalar
product \eref{eq:scalar-product} defines different monomials as pairwise
orthogonal. So the set of all possible monomials are an orthonormal basis of
the super Hilbert space of operators. 

The scalar product \eref{eq:scalar-product} implies the norm of an operator
$O$ as $ \|O\|^2 :=
\left\langle O , O \right\rangle $. We again minimize $\bra{0} \left[F(l),H(l)
\right] \ket{0}$, but with the constraint $\| F(l) \|^2 = \textrm{const}$.
Thus we calculate the variation
\begin{equation}
  \label{eq:var_second}
  \delta \bigg\{ \bra{0} \left[F(l), H(l) \right] \ket{0} +
  \lambda \left\|F(l) \right\|^2 \bigg\} = 0.
\end{equation}

The operators $H(l)$ and
$F(l)$ are expanded in second quantization 
\numparts
  \begin{equation}
    \label{eq:second_a}
    H(l) = \sum_{\left\{\mathbf{i},\mathbf{j}\right\}} 
h^{\mathbf{i}}_{\mathbf{j}}(l)  M^{\mathbf{i}}_{\mathbf{j}} 
  \end{equation}
  and
  \begin{equation}
    \label{eq:second_b}
    F(l) = \sum_{\left\{\mathbf{i},\mathbf{j}\right\}} 
    f^{\mathbf{i}}_{\mathbf{j}}(l)    M^{\mathbf{i}}_{\mathbf{j}} 
  \end{equation}
\endnumparts
with the $l$-dependent prefactors $\left\{ h^{\mathbf{i}}_{\mathbf{j}}(l)
\right\}$ and $\left\{ f^{\mathbf{i}}_{\mathbf{j}}(l) \right\}$. Here the
bold indices $\mathbf{i}$ and $\mathbf{j}$ are sets of indices,
e.g. $\mathbf{i}=\left\{i_1,\ldots,i_{N_\mathbf{i}} \right\}$. 
Upper indices stand for creation operators and lower indices for annihilation 
operators. So $M^{\mathbf{i}}_{\mathbf{j}}$ is short hand for the monomial
\begin{equation}
  M^{\mathbf{i}}_{\mathbf{j}} = e^{\dag}_{i_1} \cdots 
e^{\dag}_{i_{N_{\mathbf{i}}}}
  e^{\phantom{\dag}}_{j_1} \cdots e^{\phantom{\dag}}_{j_{N_{\mathbf{j}}}}.
\end{equation}
The sums $ \sum_{\left\{\mathbf{i},\mathbf{j}\right\}}$ in
\eref{eq:second_a} and \eref{eq:second_b} run over
all possible ordered sets $\mathbf{i}$ and $\mathbf{j}$ so that a 
unique expansion in monomials $M^{\mathbf{i}}_{\mathbf{j}}$ is achieved.

Based on \eref{eq:second_a} and \eref{eq:second_b} the right hand side of \eref{eq:var_second} 
to be varied has two additive contributions. The first one reads
\begin{eqnarray}
  \eqalign{
    \bra{0} \left[F(l),H(l)\right] \ket{0} &= \bra{0} F(l) H(l) - H(l) 
    F(l) \ket{0} \\
    &= \sum_{\left\{\mathbf{i}\right\}} \left( 
    f_{\mathbf{i}}^{\emptyset}(l) h^{\mathbf{i}}_{\emptyset}(l) -
    h_{\mathbf{i}}^{\emptyset}(l) f^{\mathbf{i}}_{\emptyset}(l)
    \right),} 
\end{eqnarray}
where the empty set $\emptyset$ stands for
the lack of non-trivial operators, in particular, a prefactor
$f_{\mathbf{i}}^{\emptyset}(l)$ belongs to a term that only
contains annihilation operators. We exploit the fact that only 
creation operators yield non-vanishing
results if applied to $\ket{0}$.  Conversely, only annihilation operators yield
non-vanishing bra states if placed right to $\bra{0}$. 

The second contribution reads
\begin{equation}
  \lambda \left( \| F(l) \|^2 \right) = \lambda \left(
    \sum_{\left\{ \mathbf{i}, \mathbf{j} \right\} } \left|
      f^{\mathbf{i}}_{\mathbf{j}}(l) \right|^2 \right).
\end{equation}
Making the variation with respect to $f^{\mathbf{i}}_{\mathbf{j}}(l)$ vanish
leads to 
\begin{equation}
  f^{\mathbf{i}}_{\mathbf{j}}(l) = \frac{1}{2\lambda} 
 \left(h^{\mathbf{i}}_{\emptyset}(l) \delta_{\mathbf{j},
    \emptyset} - \delta_{\mathbf{i},\emptyset} h_{\mathbf{j}}^{\emptyset}(l) 
 \right).
\end{equation}

This generator solely contains monomials which are only composed of creation
operators or only of annihilation operators. If we set $\lambda =
{1}/{2}$ we obtain exactly the generator $F_{\textrm{gs}}(l)$ we
conjectured in \eref{eq:eta_gs}. Note that the above derivation
holds for all kinds of operators in second quantization, including bosons,
hard-core bosons, fermions and hard-core fermions.
This terminates the derivation of $F_{\textrm{gs}}(l)$ and its properties.

In this paper, we only consider the case where the generator 
$F_{\textrm{gs}}(l)$ separates only the vacuum state from all other states. 
But we want to mention that it is also possible to generalize the generator
$F_{\textrm{gs}}(l)$ to the case where the vacuum state $\ket{0}$ is replaced 
by a statistical operator which defines a certain subspace, i.e. a
reference ensemble. In this case the
generator $F_{\textrm{gs}}(l)$ induces an effective model on the
reference subspace, which is separated from all other states.
A well-known example is the derivation of the Heisenberg model or the
$t$-$J$ model from the Hubbard model.
This generalization  works very much in the same way as it was done for the 
generator $F_{\textrm{pc}}(l)$ 
before \cite{RMHU04,Reischl2006,Lorscheid2007,Hamerla2009}.

\subsection{Other similar generators}
\label{subsec:gs_1p}
\begin{figure}[ht]
  \centering
  \includegraphics[width=0.65\columnwidth]{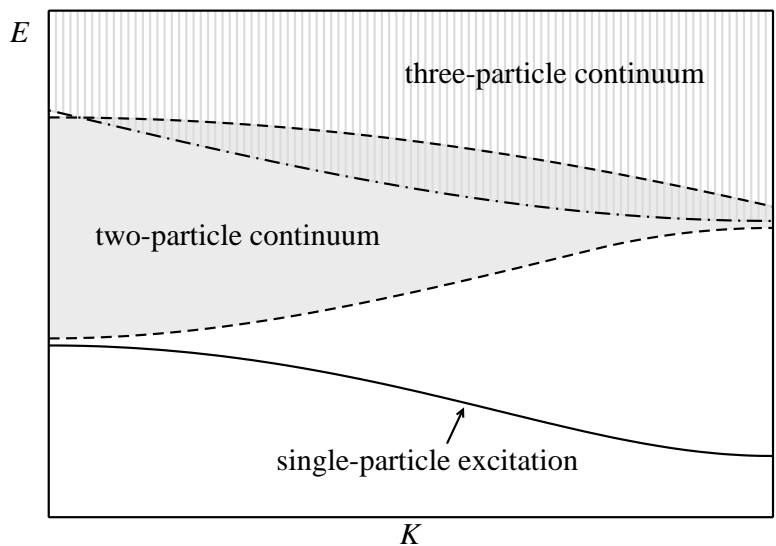}
  \caption{Overlap of the two- and three-particle continua.}
  \label{fig:overlap2}
\end{figure}

Besides the two choices of a generator considered so far 
($F_{\textrm{pc}}(l)$ in \eref{eq:eta_pc} and $F_{\textrm{gs}}(l)$ in
\eref{eq:eta_gs}) there also exist other possibilities. For
example, one can also include all terms to the generator  
$F_{\textrm{pc}}(l)$ which are connected to the one-particle subspace 
\begin{equation}
   \label{eq:eta_gs_1p}
   F_{\textrm{gs,1p}}(l)=\sum_{i>0}^N \left(H_{0}^{i}(l) -
     H_{i}^{0}(l)\right) + \sum_{i>1}^N \left(H_{1}^{i}(l) -
     H_{i}^{1}(l)\right).
\end{equation}
Since this generator also separates the one-particle subspace from all
subspaces with two and more particles, it is
not an ideal choice to describe (quasi)particle decay.
It suffers from the same caveats as $F_{\textrm{pc}}(l)$. 
But this generator can be the optimal
choice if the (quasi)particles
have an infinite lifetime, while the higher particle subspaces are overlapping
in energy (cf.\ \fref{fig:overlap2}).
In \fref{fig:comparison} the structure of the corresponding Hamiltonian
$H(l)$ is  schematically illustrated during the flow.

\subsection{Common properties}
\label{subsec:properties}
Although different generators produce different CUTs and therefore lead to 
different effective models it
happens that they transform certain subspaces in exactly the same way.
For example, it can be proven (see \sref{app:sub:gs}) that all
generators considered in \sref{subsec:eta_pc}, \sref{subsec:gs} and
\sref{subsec:gs_1p} transform the vacuum state $\ket{0(l)}$ equally. 
This is a consequence of the fact that for all these generators the
matrix elements from and to the ground state $F_{i,0}(l)$ and
$F_{0,i}(l)$, respectively, are defined in the same way as long as
the flow equation is treated exactly without any truncation.

\begin{figure}[ht]
  \centering
  \includegraphics[width=0.65\columnwidth]{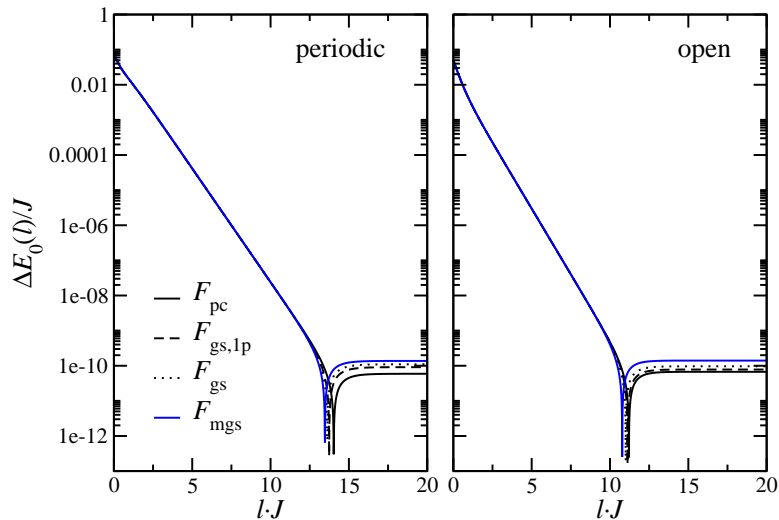}
  \caption{Evolution of $\Delta E_0(l):=\left|E_0(l)-E_{\textrm{exact}}\right|$ 
    during the flow for an antiferromagnetic spin-${1}/{2}$
    Heisenberg chain with $10$ spins and exchange coupling
    $J=J_{\perp}$ and $x=0$, $y=1$ (cf.\ \eref{eq:H_aasl_a}-\eref{eq:H_aasl_d}). All
    calculations started from the 
    dimerized phase. The left panel shows the results for periodic 
    boundary conditions; the right panel shows the results for open 
    boundary conditions.}
  \label{fig:gs_chain}
\end{figure}

In \fref{fig:gs_chain}, we show numerical data verifying the equivalent
transformation of the vacuum state $\ket{0(l)}$ by different generators. 
The $l$-dependence of the difference 
$\Delta E_0(l):=\left|E_0(l)-E_{\textrm{exact}}\right|$ 
between the vacuum expectation value $E_0(l):=\bra{0}H(l) \ket{0}$ 
 and the exact ground state energy $E_{\textrm{exact}}$ is plotted for the 
 different generators. The
system under study is an antiferromagnetic spin-${1}/{2}$
Heisenberg chain with $10$ spins and exchange coupling $J$. The starting point 
for all calculations
is the ground state and the local triplons of the completely dimerized phase 
(cf.\ \sref{sec:model}). We considered periodic and
open boundary conditions. \Fref{fig:gs_chain} shows clearly that all
considered generators transform the vacuum state $\ket{0(l)}$ in the same way.
The features beyond $l\approx 12/J_\perp$ stem from
numerical inaccuracies occurring at $\Delta E \approx 10^{-10}J$.
These inaccuracies are shown here to illustrate where and how
numerical errors make themselves felt.

Similarly, one can prove that the generator $F_{\textrm{pc}}(l)$ and the 
generator $F_{\textrm{gs,1p}}$ transform all one-particle states
identically (see \sref{app:sub:1p}).

\section{Model: Asymmetric antiferromagnetic spin-${1}/{2}$ Heisenberg
ladder}
\label{sec:model}
\begin{figure}
  \centering
  \includegraphics[width=0.65\columnwidth]{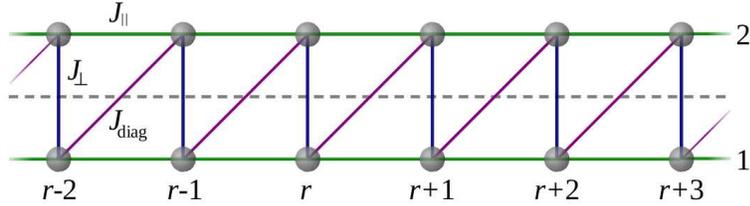}
  \caption{\label{fig:ladder_asym}
    Schematic representation of the asymmetric spin ladder. The additional 
    diagonal interaction $J_{\textrm{diag}}$ breaks the reflection symmetry and 
    hence induces a hybridization between the
    one-quasiparticle states and the two-quasiparticle
    continuum. Thus it is an ideal model to study quasiparticle breakdown.}
\end{figure}

The Hamiltonian for the asymmetric antiferromagnetic 
($J_\parallel,J_\perp,J_{\textrm{diag}}>0$) spin-${1}/{2}$ Heisenberg
ladder reads
\numparts
\begin{equation}
  \label{eq:H_aasl_a}
  H = J_{\perp}\left(H_\perp + x H_\parallel + y
  H_{\textrm{diag}}\right)
\end{equation}
with
\begin{eqnarray}
  \label{eq:H_aasl_b} H_\perp &=& \sum_r \boldsymbol{S}_{1,r} \boldsymbol{S}_{2,r} \\
  \label{eq:H_aasl_c} H_\parallel &=& \sum_r \left(\boldsymbol{S}_{1,r}
  \boldsymbol{S}_{1,r+1} + \boldsymbol{S}_{2,r}
  \boldsymbol{S}_{2,r+1}\right) \\
  \label{eq:H_aasl_d} H_\textrm{diag} &=& \sum_r  \boldsymbol{S}_{1,r} \boldsymbol{S}_{2,r+1},
\end{eqnarray}
\endnumparts
where the first subscript $1,2$ denotes the leg and $r$ the rung (see
 \fref{fig:ladder_asym}). The parameter $x$ is given by 
$ x := {J_\parallel}/{J_\perp}$ and the
parameter $y$ by  $y := {J_\textrm{diag}}/{J_\perp}$.

\begin{figure}
  \centering
  \includegraphics[width=0.65\columnwidth]{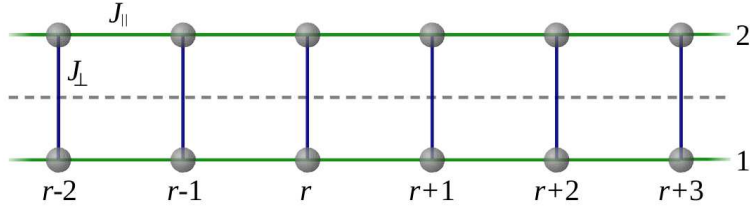}
  \caption{\label{fig:ladder}
    Schematic representation of a spin ladder. Circles indicate
    spins with $S={1}/{2}$. Solid lines stand for couplings. 
    The dashed line indicates the axis of reflection symmetry.}
\end{figure}

This Hamiltonian contains some frequently discussed models. For
example, for $x=0$ and $y=1$ the Hamiltonian \eref{eq:H_aasl_a}-\eref{eq:H_aasl_d}
describes the exactly solvable isotropic antiferromagnetic
spin-${1}/{2}$ Heisenberg
chain \cite{Bet31,Hul38,CP62,YY66_1,YY66_2,FT81,Bax82}. 
In the broad field of spin systems without magnetic long-range order the limit 
of the symmetric spin-${1}/{2}$ Heisenberg
ladder \cite{BDRS93,DR96,SK98,DS98,BMMNU99,JB00,trebs00,KSGU01,ZHSTM01,SKU01,HS02,KSU04twoleg,SU05} (see \fref{fig:ladder}) with $y=0$ is very popular as one-dimensional
example of a valence-bond solid. Besides the theoretical interest in $H$
in \eref{eq:H_aasl_a}-\eref{eq:H_aasl_d}, there is a large number of compounds which can be 
described by spin ladders (see e.g. \cite{K_etal95,Schwenk_etal96,EAT96,KTKK97,HRBT98,SS99,MKEBM00,KIIU02,Grueninger_etal02,Notbohm_etal07}
or for an overview \cite{Johnston_etal00}). 
Special interest has been raised by the realization of coupled spin 
ladders in the stripe
phases of cuprate superconducters \cite{TSANU95,VU04,USG04}. 
Also the experimental evidence for
superconductivity in Sr$_{0.4}$Ca$_{13.6}$Cu$_{24}$O$_{41}$ under
pressure \cite{U_etal96} contributed to the interest in the 
spin-${1}/{2}$ Heisenberg ladder and its extended versions. 
For the case $x<y$ the Hamiltonian \eref{eq:H_aasl_a}-\eref{eq:H_aasl_d} is usually
denoted as dimerized and frustrated spin-${1}/{2}$ Heisenberg
chain (see \cite{SKU04} and references therein).

In the following, we study the two parameter sets
$x=0.5$, $y=0$ and $x=0.5$, $y=0.1$ in the thermodynamic limit as
generic examples.
Since in both cases the relation $x>y$ is fulfilled we call the system an
asymmetric  ladder instead of a dimerized and frustrated chain.
A more comprehensive investigation of the dependences of the quasiparticle 
decay on the model parameter is left to ongoing research. The scope
of this  section and the next one is to illustrate the general considerations
concerning the method by a concrete example.

The low energy spectrum for $x=0.5$ and $y=0$ is well studied by several
methods (see e.g. \cite{SK98} and \cite{trebs00}) 
including methods based on CUTs \cite{KSU04}. Therefore, it is a
perfect starting point to discuss the more sophisticated case with
$x=0.5$ and $y=0.1$. The additional diagonal interaction $y$ makes the whole
situation conceptionally more difficult because it breaks a symmetry. While for
 $y=0$ the model is symmetric under reflection (see \fref{fig:ladder})
an arbitrary small value $y\neq 0$ breaks this
reflection symmetry (see \fref{fig:ladder_asym}). 

The crucial point is that the reflection symmetry of the symmetric ladder is
responsible for the infinite lifetime of the triplons, which are the
$S=1$ elementary magnetic excitations of an antiferromagnetic system
without long-range order \cite{SU03}. Breaking this symmetry creates processes which enable the triplons to
decay into two-triplon states. Therefore, the asymmetric spin-${1}/{2}$ Heisenberg
ladder is an ideal model to analyze quasiparticles with finite lifetime
and to illustrate our previous theoretical considerations concerning
the choice of an adaptive generator quantitatively. 

To define an appropriate starting point for the CUTs we use the bond operator
representation \cite{Chubukov89,SB90}. Each rung $r$ of the ladder considered separately
has a four dimensional Hilbert space. A possible eigenbasis of the
local operator $\boldsymbol{S}_{1,r} \boldsymbol{S}_{2,r}$ is given by
the singlet state
\numparts
\begin{equation}
  \ket{s}_r := \frac{1}{\sqrt{2}} \left( \ket{\uparrow \downarrow}-
  \ket{\downarrow \uparrow} \right)_r
\end{equation}
and the three triplet states
\begin{eqnarray}
  t_{x,r}^{\dag}\ket{s}_r  := \ket{t_{x}}_r =& \frac{-1}{\sqrt{2}} \left( \ket{\uparrow \uparrow}-
  \ket{\downarrow \downarrow} \right)_r \\     
  t_{y,r}^{\dag}\ket{s}_r  :=  \ket{t_{y}}_r =& \frac{\rmi}{\sqrt{2}} \left( \ket{\uparrow \uparrow}+
  \ket{\downarrow \downarrow} \right)_r \\     
  t_{z,r}^{\dag}\ket{s}_r  := \ket{t_{z}}_r =& \frac{1}{\sqrt{2}} \left( \ket{\uparrow \downarrow}+ \ket{\downarrow \uparrow} \right)_r.
\end{eqnarray}
\endnumparts
Without any interactions along the ladder ($x=0$, $y=0$) the ground state of
the system is the product state of the rung singlets 
\begin{equation}
  \ket{0} := \prod_r \ket{s}_r.
\end{equation}
This reference state shall be the vacuum state of the
system. Excitations on a rung $r$ are created by the local operators
$t_{x,r}^{\dag}$, $t_{y,r}^{\dag}$ and $t_{z,r}^{\dag}$. These
operators create a triplet on the rung $r$ and
satisfy the hard-core boson commutation relations
\begin{equation}
  \left[ t_{\alpha,r}^{\phantom{\dag}},t_{\beta,s}^{\dag} \right] = \delta_{r,s} \left(
    \delta_{\alpha,\beta} - t_{\beta,r}^{\dag}t_{\alpha,r}^{\phantom{\dag}} -
    \delta_{\alpha,\beta} \sum_{\gamma}\left(t_{\gamma,r}^{\dag}t_{\gamma,r}^{\phantom{\dag}} \right) \right),
\end{equation}
where $ t_{\alpha,r}$ ($\alpha=x,y,z$) annihilate such a triplet.
We consider all the excited states, which can be continuously connected
to the local triplets, to be the elementary magnetic excitations. They are
called triplons \cite{SU03,SU05}.

In the bond operator representation the Hamiltonian \eref{eq:H_aasl_a}-\eref{eq:H_aasl_d}
is given in the notation of \eref{eq:hamiltonian} by
\numparts
  \begin{equation}
    \label{eq:H_bond_a}
    \fl  H=J_{\perp}\left(H_0^0+H_1^1+H_2^2+H_0^2+H_2^0+H_1^2+H_2^1\right)
  \end{equation}
  with
  \begin{eqnarray}    
    \fl    H_0^0 =-\sum_r \frac{3}{4}\\
    \eqalign{
      \fl H_1^1 =\sum_r t^{\dag}_{\alpha,r}
      t^{\phantom{\dag}}_{\alpha,r} +\left(\frac{1}{2}x-\frac{1}{4}y\right)\sum_r \sum_\alpha
      \left(t^{\dag}_{\alpha,r} t^{\phantom{\dag}}_{\alpha,r+1} +
      t^{{\dag}}_{\alpha,r+1} t^{\phantom{\dag}}_{\alpha,r} \right)
    }\\
     \eqalign{
       \fl H_2^2 =\left(\frac{1}{2}x+\frac{1}{4}y\right) 
    \sum_r \sum_{\alpha \neq \beta}  t^{\dag}_{\alpha,r} 
    t^{{\dag}}_{\beta,r+1} t^{\phantom{\dag}}_{\beta,r} 
    t^{\phantom{\dag}}_{\alpha,r+1} \\ 
      -\left(\frac{1}{2}x+\frac{1}{4}y\right) \sum_r \sum_{\alpha \neq \beta} 
      t^{\dag}_{\alpha,r} t^{{\dag}}_{\alpha,r+1} t^{\phantom{\dag}}_{\beta,r} 
      t^{\phantom{\dag}}_{\beta,r+1} 
    }\\
    \fl H_0^2 = \left(\frac{1}{2}x-\frac{1}{4}y\right) \sum_r \sum_\alpha  t^{\dag}_{\alpha,r} t^{{\dag}}_{\alpha,r+1}\\
    \fl H_2^0 = \left( H_0^2 \right)^{\dag}\\
    \label{eq:Hg}
    \fl H_1^2 =- \frac{\rmi}{4}y \sum_r \sum_{\alpha, \beta, \gamma} \varepsilon_{\alpha \beta \gamma} \left( t^{\dag}_{\alpha,r} t^{{\dag}}_{\beta,r+1} \left( t^{\phantom{\dag}}_{\gamma,r} +  t^{\phantom{\dag}}_{\gamma,r+1} \right) \right)\\
    \label{eq:Hh}
    \fl H_2^1=\left( H_1^2 \right)^{\dag}.
  \end{eqnarray} 
\endnumparts

This representation of the Hamiltonian of the asymmetric antiferromagnetic 
spin-${1}/{2}$ Heisenberg ladder is used as the starting point for 
the CUTs. For $y=0$ the terms \eref{eq:Hg} and \eref{eq:Hh} vanish whereby
decay processes of one triplon into two are prevented.

Evaluating the commutator $\left[F(l),H(l)\right]$ appearing on the
right hand side of the flow equation \eref{eq:flow_equation} generates
also terms which do not appear in the initial Hamiltonian
$H(0)=H$. These terms must be added to the Hamiltonian with a
coefficient equal to zero at $l=0$. They must also be
considered in the generator. Then the commutator
$\left[F(l),H(l)\right]$ generates even more terms which have to be
taken into account. For a finite-dimensional Hilbert space this procedure 
comes to an end  because the maximal number of terms is restricted. 
Such unrestricted calculations were performed to compute the results for the 
ground-state energy of
the finite Heisenberg chain presented in \sref{subsec:properties}.

For large systems such an unrestricted approach is not
possible due to the proliferating number of terms. Especially in the
thermodynamic limit, one has to deal with an infinite number of
terms. Hence it is not possible to obtain a closed set of differential 
equations. Thus in practice one has to decide which terms are important to 
describe the underlying physics properly and which terms
can be neglected. 

One established truncation scheme is to use a
perturbative approach \cite{KU00,KSU03} (p-CUT) which is based on the
generator $F_{\textrm{pc}}(l)$. But since we intend to describe the decay of
quasiparticles so that variations of the generator
$F_{\textrm{pc}}(l)$ (see \sref{sec:method}) have to be used,
we choose the  \emph{self-similar} approach (s-CUT). 
But there is no fundamental reason why the adapted generator cannot
be implemented perturbatively as well.

The s-CUT was used in many previous
applications of the CUTs (for an overview see \cite{Kehrein2006} and
 references therein) among them the original work by Wegner \cite{weg94}.
The whole transformation takes place in the coefficients of the terms in the 
Hamiltonian  which motivates the naming ``self-similar''.
This approach can straightforwardly be implemented for various
generators.

In the present paper, we apply the following truncation scheme, which
is based on the finite correlation length of one-dimensional systems
with a finite gap.
The truncation scheme is based on the locality of the term
which is justified for systems with finite correlation length.
It is described in detail in the following \sref{subsec:truncation}.

All in all we proceed in the actual calculations as follows:
\begin{itemize}
\item Define a truncation scheme which restricts the maximal number of terms.
\item Set up the flow equation by
  calculating the commutator $\left[F(l),H(l)\right]$. Only
  those terms are considered which fit the truncation scheme.
\item Solve the flow equation numerically.
\end{itemize}

For calculations in the thermodynamic limit, it is necessary to make use of the
translational invariance of the Hamiltonian \eref{eq:H_bond_a}-\eref{eq:Hh}. The translation
symmetry ensures that terms which describe identical processes except
for a shift along the
ladder have the same coefficient. Consequently, it is sufficient to track only 
one representative of this symmetry group. This procedure is also possible for 
all others symmetries of the Hamiltonian \eref{eq:H_bond_a}-\eref{eq:Hh}, e.g., the spin
symmetry and the rotational symmetry by $\pi$ of the ladder. Using
representatives of the underlying symmetries reduces the number of 
coefficients appearing in the flow equation  \eref{eq:flow_equation}
significantly for a given truncation scheme. Thus more extended
truncation schemes considering more processes become feasible. 

\subsection{Truncation scheme} 
\label{subsec:truncation}
A truncation scheme is necessary to limit the number of terms so that
a closed set of differential equations is achieved. In the present paper, we 
are dealing with terms in real-space and the truncation scheme is based on the
locality of the terms. We first define a measure for the
locality of a term, which we call the extension $d$. The extension $d$ of
a term is defined by the distance between the rightmost to the
leftmost rung on which the monomial acts in a nontrivial way. For example, the 
term $t^{\dag}_{\alpha,r} 
t^{{\dag}}_{\beta,r+1} t^{\phantom{\dag}}_{\gamma,r+4}$
has an extension $d=4$. Second, we define the truncation scheme by
choosing a maximal extension $d_{\textrm{max}}$ discarding all terms with a
larger extension ($d>d_{\textrm{max}}$). 

It turns out that it is appropriate to define
not only one maximal extension for all terms but to keep
terms with a different number of annihilation or creation operators up to
different maximal extensions \cite{Reischl2006}.
Accordingly, terms with $n$ annihilation or creation operators in
total are required to have an
extension $d_{n}$ or less to be kept in the flow equation. As a second
truncation criterion we admit only terms which create or annihilate not
more than $N$ (quasi)particles. Thus the total truncation scheme is defined by
the value of $N$ and the set of extensions
$\mathbf{d}=\left(d_2,\ldots,d_{2N}\right)$. 
Note that due to the conservation of spin no single triplon operators can
occur in the Hamiltonian. So no $d_1$ needs to be denoted. In addition, the
translational invariance of the Hamiltonian  makes $d_1$
superfluous. In this case only six different monomials exist which
act on one rung only, in particular, $t^{\dag}_{\alpha,r}$ and
$t^{\phantom{\dag}}_{\alpha,r}$ with $\alpha=x,y,z$.
For the
symmetric ladder ($y=0$) no terms occur which consist of an odd
number of operators. Therefore, we do not need to define maximum
extensions $d_3,d_5,\ldots$ . In the notation of the set of extensions 
$\mathbf{d}$ we replace such superfluous extensions by a dot, e.g. 
$\mathbf{d}=(8,.,6,.,4)$.

It is worthwhile to emphasize that this truncation scheme does not turn our
approach to a calculation on a finite cluster. It is a self-similar
calculation strictly in the thermodynamic limit. We only truncate the range of 
the interactions in real space, but not the Hilbert space.

\section{Results}
\label{sec:results}
Here we present results for the symmetric ladder with $x=0.5$
and $y=0$ and for the asymmetric ladder with $x=0.5$ and $y=0.1$. These two
parameter sets are chosen to illustrate the differences between
systems with stable quasiparticles and systems with unstable
quasiparticles which exhibit
a finite lifetime. In particular, we confirm our previous statements
concerning the properties of the different generators (cf.\ \sref{sec:method}).

Firstly we show that a rearrangement of the states of the
Hilbert space, i.e., a continuous re-labelling (for simple examples
see \cite{DU04}), reduces the speed of convergence (see \sref{sec:convergence}). Therefore, generators which
avoid such a rearrangement induce a considerably faster convergence. Secondly,
we discuss the low energy spectrum for the symmetric and for the
asymmetric ladder (see \sref{sec:low_energy_spectrum}). If decay is 
possible the generator $F_{\textrm{gs,1p}}(l)$ and the
generator $F_{\textrm{pc}}(l)$ indeed
tend to interpret the energetically lowest states above the
ground state as the elementary excitations (as stated before
in \sref{sec:method}). This can be avoided by using the generator
$F_{\textrm{gs}}(l)$. Unfortunately, for this generator a
simple calculation in the one-particle subspace is not sufficient to
arrive at reliable results for the true one-triplon dispersion. This is a
consequence of the fact that in the effective Hamiltonian induced by 
$F_{\textrm{gs}}(l)$ the one-particle
subspace still couples to higher particle subspaces (see \fref{fig:comparison}). To obtain reliable results for the single
triplon dispersion  we include states which consist of up to three particles
in our calculations (see \sref{sec:spectral_density}). 
Especially, we show
results for the zero temperature spectral density in which the
(quasi)particle decay is manifest as a Lorentzian resonance of finite width.
\subsection{Convergence}
\label{sec:convergence}
To quantify the speed of convergence of the flow equation for different
generators we introduce the \emph{residual off-diagonality} 
(ROD) \cite{RMHU04,Reischl2006}. The
ROD is defined as the square root of the sum of the moduli squared of all
coefficients that contribute to the considered generator. Using the
notation of \eref{eq:second_b} the ROD is given by
\begin{eqnarray}
  \label{eq:rod}
  \textrm{ROD}(l) = \sqrt{\sum_{\left\{\mathbf{i},\mathbf{j}\right\}} \left|
f^{\mathbf{i}}_{\mathbf{j}}(l)\right|^2}
\end{eqnarray}  
where the range of the sum $\sum_{\left\{\mathbf{i},\mathbf{j}\right\}}$
depends on the choice of the generator.
Note that only one representative of the translational symmetry group is
included. Otherwise the ROD would grow proportional to the system
size. In addition, ROD$^i_j$
denotes the square root of the sum of the moduli squared of all coefficients 
belonging to terms with $i$ creation and $j$ annihilation operators or to 
their  Hermitian conjugate terms.

\begin{figure*}[ht]
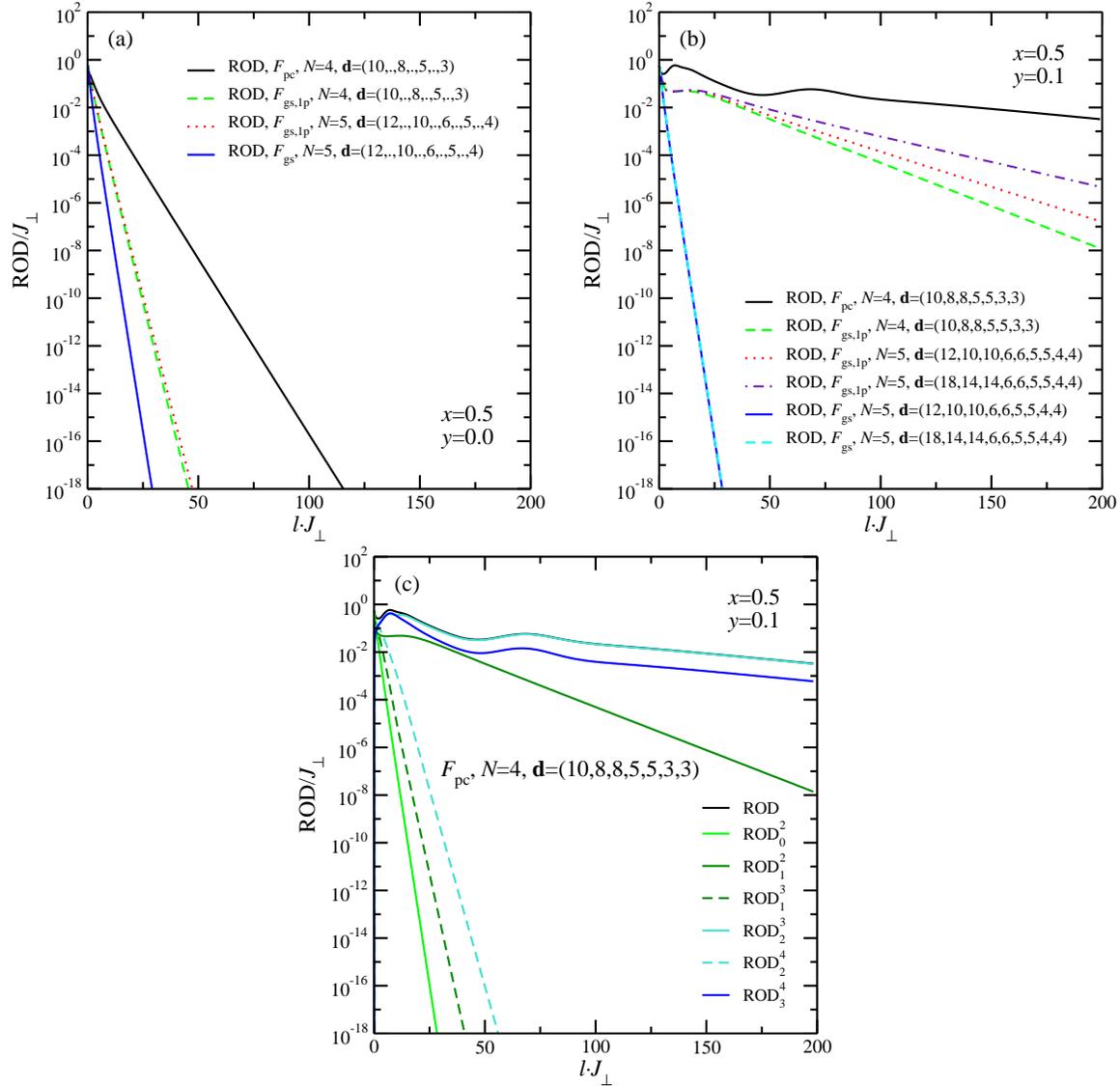

  \begin{minipage}{0.495\textwidth}
    \includegraphics[width=0.95\textwidth]{figure7a}
  \end{minipage}
  \begin{minipage}{0.495\textwidth}
    \includegraphics[width=0.95\textwidth]{figure7b}
  \end{minipage}\\
  \centering
  \begin{minipage}{0.495\textwidth}
    \includegraphics[width=0.95\textwidth]{figure7c}
  \end{minipage}
  \caption{\label{fig:rod}
    Convergence of the flow equations. Panel (a) shows the
    evolution of the ROD during the flow for
    different generators and different truncation schemes for the symmetric
    ladder ($x=0.5$ and $y=0$). In all cases the ROD decreases strictly
    monotonically. Panel (b) shows the evolution of the ROD for
    different generators and different truncation schemes for the 
    asymmetric ladder ($x=0.5$ and $y=0.1$). The RODs of the generator 
    $F_{\textrm{gs,1p}}(l)$ and the
    generator $F_{\textrm{pc}}(l)$ increase temporarily during the flow.
    This indicates a significant rearrangement of the states in
    the Hilbert space. Panel (c) shows the ROD of the generator 
    $F_{\textrm{pc}}(l)$ split in the parts ROD$^i_j$ for the asymmetric
    ladder ($x=0.5$ and $y=0.1$). The main contributions to the total ROD
    is due to  ROD$^i_j$ with $|i-j|=1$.}
\end{figure*}

\Fref{fig:rod}a shows the evolution of the ROD during the flow for
different generators and different truncation schemes for $x=0.5$ and $y=0$.
For all generators the RODs decrease strictly monotonically. The ROD of
the generator $F_{\textrm{gs}}(l)$ decreases faster than the ROD of the 
generator $F_{\textrm{gs,1p}}(l)$. This is a consequence of the fact that the
generator $F_{\textrm{gs,1p}}(l)$ contains more coefficients than the
generator $F_{\textrm{gs}}(l)$. The convergence of these additional
coefficients is slower because they connect states which differ less
in their eigenenergies (cf.\ \eref{eq:mku_flow_5}), e.g. the energy
gap between one- and three-triplon states is smaller than the energy
gap between the vacuum state and the two-triplon states. This also explains 
why the generator $F_{\textrm{gs,1p}}(l)$ converges faster than the generator
$F_{\textrm{pc}}(l)$. 

The convergence behavior clearly changes if one includes the diagonal
interaction, even if $y$ is small ($y=0.1$). In \fref{fig:rod}b the ROD 
during the flow for different generators and different truncation schemes for 
$x=0.5$ and $y=0.1$ is depicted. Only the ROD of the generator 
$F_{\textrm{gs}}(l)$ decreases strictly
monotonically while the RODs of the generator $F_{\textrm{gs,1p}}(l)$ and the
generator $F_{\textrm{pc}}(l)$ increase temporarily during the flow. These
increases indicate a rearrangement in the Hilbert space,
cf. \cite{DU04} for simple examples. If all
eigenstates were ordered in such a way that states with more triplons had
higher eigenenergies the ROD would decrease exponentially (cf.\ \eref{eq:mku_flow_4}). These rearrangements
affect the results for the one-triplon dispersion as we illustrate in the \sref{sec:low_energy_spectrum}. 

\Fref{fig:rod}c shows the ROD of the generator $F_{\textrm{pc}}(l)$
split in the partial RODs ROD$^i_j$ defined above for $x=0.5$ and
$y=0.1$. Clearly, the contributions ROD$^i_j$ of the
ROD changing the number of triplons only by one ($|i-j|=1$) provide the
main contributions to the total ROD, although the corresponding
initial couplings are proportional to $y$, which is small ($y=0.1$).
From this we infer that the convergence of the flow equation is mainly
influenced by terms which induce a rearrangement of the Hilbert
space if they are to be eliminated by the CUT. It is less important
whether the corresponding coupling parameter is large or not. This is
an important property of the CUTs which distinguishes them from
conventional diagrammatic perturbation theories.

In summary, we state that a rearrangement of the states of the Hilbert space 
reduces the speed of convergence. Omitting the corresponding terms from the
generator stabilizes the flow in the sense that the convergence is
monotonic and robust. Hence, especially the generator
$F_{\textrm{gs}}(l)$ yields a very fast converging and robust flow.
We point out that a fast convergence is advantageous because it
minimizes the interval in $l$ during which significant terms are truncated.
Hence as a rule of thumb, the faster the convergence, the smaller
are the truncation errors.

\subsection{Low energy spectrum}
\label{sec:low_energy_spectrum}
\begin{figure*}[ht]
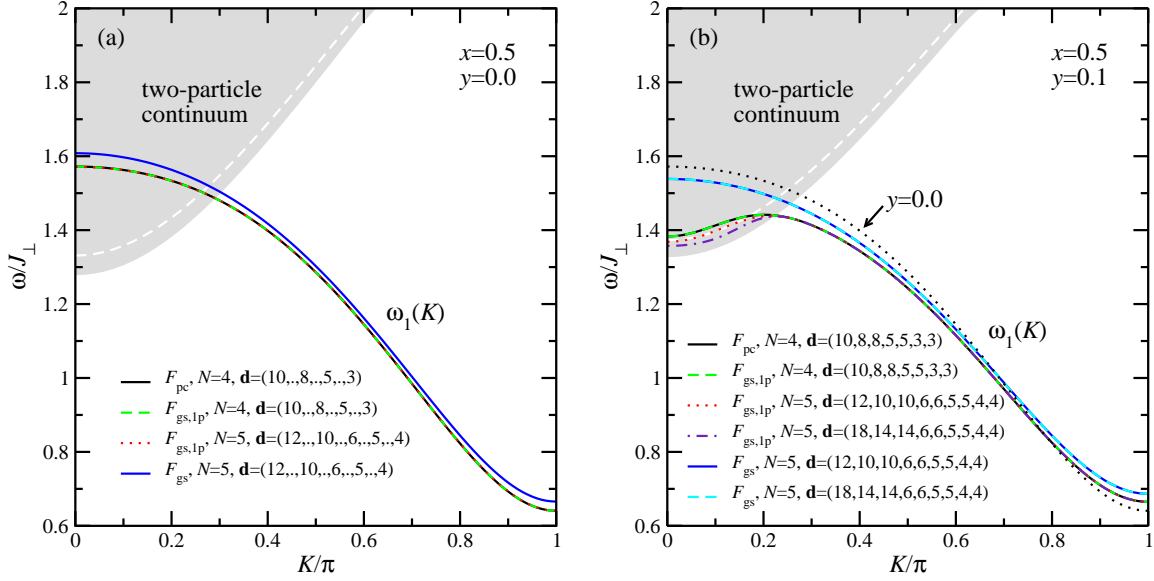

  \begin{minipage}{0.495\textwidth}
    \centering
    \includegraphics[width=0.95\textwidth]{figure8a}
  \end{minipage}
  \begin{minipage}{0.495\textwidth}
    \centering
    \includegraphics[width=0.95\textwidth]{figure8b}
  \end{minipage}
  \caption{\label{fig:disp}
    Low energy spectrum of the symmetric 
    and asymmetric spin-${1}/{2}$  Heisenberg ladder. Panel (a)
    shows results for the one-triplon dispersion $\omega_1(K)$
    for the symmetric ladder with $x=0.5$ and
    $y=0$. Results for different generators and different truncation
    schemes are depicted. Additionally, the lower part of the
    two-particle continuum is shown (grey area). The dashed white line 
    represents an approximation of
    the lower edge of the two-particle continuum obtained by the
    approximate one-triplon dispersion $\omega_1(K)$ in the case of the
    generator $F_{\textrm{gs}}(l)$. Panel (b) shows the corresponding
    quantities for the asymmetric ladder with  $x=0.5$ and $y=0.1$.}
\end{figure*}

Here we discuss the low energy
spectrum of the effective Hamiltonian $H_{\textrm{eff}}$. We always
stopped the CUT at $l=200/J_\perp$. At this value the remaining effect on the
one-particle subspace is small in all cases  (cf.\ \fref{fig:rod}) so 
that a  further integration of the flow equation would not change the results 
for the  one-triplon dispersion as shown in \fref{fig:disp}.  
  
To calculate the one-triplon dispersion $\omega_1(K)$ of the effective
model we define the Fourier-transformed one-particle states
\begin{equation}
  \ket{K,\alpha} := \frac{1}{\sqrt{\mathcal{N}}} \sum_r \e^{\rmi K r} 
\ket{r,\alpha} 
\end{equation}
with $\ket{r,\alpha}:=t^{\dag}_{\alpha,r}\ket{0}$. The action of
$H_{1}^{1}(l)$ with respect to the translational symmetry is given by \eref{eq:h_11|1} in \sref{app:analysis}. Due to the SU(2)
symmetry of the Hamiltonian $H(l)$ the hopping coefficients
$c_{11;{r}}^{\alpha',\alpha}$ obey the relation
$c_{11;{r}}^{\alpha',\alpha}= \delta_{\alpha',\alpha} c_{11;r}$. This
leads to the threefold degenerate one-triplon dispersion
\begin{equation}
  \label{eq:one_disp}
  \omega_1(K) := \bra{K,\alpha} H_{1}^{1}(\infty) \ket{K,\alpha} =
  \sum_{{r}} \e^{\rmi K {r}} c_{11;{r}} \ .
\end{equation}

The generator $F_{\textrm{pc}}(l)$ and the generator
$F_{\textrm{gs,1p}}(l)$ separate the one-particle subspace from the
other subspaces. Consequently,
for these two generators the one-triplon dispersion $\omega_1(K)$
yields eigenvalues of the effective Hamiltonian $H_{\textrm{eff}}$ in
the one-particle subspace. 
In contrast, the generator $F_{\textrm{gs}}(l)$ does not separate the 
one-particle space. Therefore, the effective Hamiltonian 
$H_{\textrm{eff}}$ still contains terms which connect the one-particle
subspace with higher particle states. In this 
case the quantity $\omega_1(K)$ only gives an approximation of the eigenvalues 
of the effective Hamiltonian $H_{\textrm{eff}}$  (cf.\ \fref{fig:subspaces}a).

In \fref{fig:disp}a the one-triplon dispersion $\omega_1(K)$ is displayed 
for $x=0.5$ and $y=0$. 
Results for all three generators 
$F_{\textrm{pc}}(l)$, $F_{\textrm{gs,1p}}(l)$ and $F_{\textrm{gs}}(l)$ and 
various truncation schemes are shown. The two generators
$F_{\textrm{pc}}(l)$ and $F_{\textrm{gs,1p}}(l)$ separating the one-particle
space yield almost the same results and barely depend
on the chosen truncation scheme. Together with the good convergence
(cf.\ \fref{fig:rod}a) this implies that the results are very reliable.
By construction, for the generator $F_{\textrm{gs}}(l)$ the quantity 
$\omega_1(K)$ as defined above yields only an approximation of the true 
one-triplon dispersion. The resulting $\omega_1(K)$ is an 
upper bound to the results obtained from the other two generators
if the truncation errors are negligible. This fact is based on 
the variational principle that a minimum in a restricted subspace is an upper 
bound to the minimum in an  unrestricted subspace.
To improve the results in this case one has to consider higher particle 
subspaces as well (cf.\ \sref{sec:spectral_density}). 

\Fref{fig:disp} also displays the lower part of the two-particle continuum
\begin{equation}
  \label{eq:2p_cont}
  \omega_2(K,Q)= \omega_1\left({K}/{2}+Q\right)+
  \omega_1\left({K}/{2}-Q\right),
\end{equation}
where $Q \in \left[ -\pi,\pi \right]$ denotes the relative
momentum. The lower band edge is given by the minimum of
$\omega_2(K,Q)$ over $Q$; the maximum yields the upper band edge, respectively.
Here we used the one-triplon dispersion $\omega_1(K)$ we obtained by 
the generator $F_{\textrm{pc}}(l)$. The additional dashed white line 
represents an approximation of
the lower edge of the two-particle continuum obtained by the
approximate one-triplon dispersion $\omega_1(K)$ in the case of the
generator $F_{\textrm{gs}}(l)$. We emphasize again that due to the reflection 
symmetry for $y=0$ (cf.\ \fref{fig:ladder}) no 
interaction exists between the one-triplon states and the
two-particle continuum.  As a result the quasiparticles are
well-defined and infinitely long-lived for the whole Brillouin zone, although 
the two-particle continuum starts below the 
one-triplon dispersion for  certain momenta $K$. In addition, this
symmetry prevents any rearrangement between the one- and
two-particle subspaces during the flow (see \sref{app:ordering}). 
This situation  changes abruptly if a diagonal interaction is switched
on, even if $y$  is very small. 

In \fref{fig:disp}b the one-triplon dispersion $\omega_1(K)$ is
displayed for $x=0.5$ and a small additional diagonal interaction $y=0.1$. 
Again results for all three generators $F_{\textrm{pc}}(l)$, 
$F_{\textrm{gs,1p}}(l)$ and $F_{\textrm{gs}}(l)$ and various truncation
schemes are shown as well as the lower part of the two-particle continuum 
$\omega_2(K,Q)$ determined from the one-triplon dispersion
obtained by the generator $F_{\textrm{pc}}(l)$. Likewise, the approximate
results for the lower edge of the two-particle continuum obtained by
the generator $F_{\textrm{gs}}(l)$ are shown. For comparison only, we also 
included the former results of the one-triplon dispersion
obtained by the generator $F_{\textrm{pc}}(l)$ for $x=0.5$ and $y=0$.

The use of the two generators
$F_{\textrm{pc}}(l)$ and $F_{\textrm{gs,1p}}(l)$ implies significantly lower
energies for the  one-triplon dispersion, see \fref{fig:disp}b,
where $\omega_1(K)$ overlaps with the two-triplon continuum.
The results strongly depend on the truncation scheme in this region.
This can be explained as follows. Since for $y\neq 0$ the one-particle
and the two-particle space are interacting with each other, the two
generators $F_{\textrm{pc}}(l)$ and $F_{\textrm{gs,1p}}(l)$ try to sort the 
eigenvalues in such a way that the eigenvalues of the 
one-triplon dispersion lie below the two-particle continuum, see \fref{fig:decay} and \sref{app:ordering}. Therefore, the one-triplon 
dispersion of the effective model $H_{\textrm{eff}}$ lies at the lower edge of 
the  two-particle continuum in the region where the one-triplon
dispersion merges with the two-particle continuum. This is not
completely achieved in practice because of the indispensable usage of a 
truncation scheme. 

We truncate the range of the decay processes in real space. This means
that the distance between the generated two triplons is limited although
the true scattering state comprises contributions up to infinite distances.
As a result, the rearrangement of the eigenvalues is only incomplete.
\Fref{fig:disp}b illustrates that increasing the range of the decay
processes (e.g. increasing $d_3$) implies that $\omega_1(K)$ approaches
the lower band edge of  $\omega_1(K,Q)$ from above more and more.

As stated before, the rearrangements of the states are unfavorable
for two reasons. (i) They imply a slow convergence which may cause growing
truncation errors. (ii) One usually defines the state with the largest
spectral weight above the ground state as the elementary excitation of the 
system and not a state with  almost no spectral weight, even if
it is lower in energy.

To avoid the rearrangement of the eigenstates, which leads to a potentially
misleading quasiparticle picture, we employ the operator $F_{\textrm{gs}}(l)$
(cf.\ \fref{fig:disp}b). As before the generator $F_{\textrm{gs}}(l)$ only
yields an approximation for
the one-triplon eigenvalues of the effective Hamiltonian 
$H_{\textrm{eff}}$. This is the case even in the region of the Brillouin zone 
where the quasiparticles are well-defined. Due to our treatment of the problem
in real space we cannot distinguish processes in different
regions in momentum space easily. To improve the results for the
one-triplon dispersion one must consider the interaction with states
which consists of more than one particle as well. This is discussed in the 
next \sref{sec:spectral_density}.

Here, we first want to show the results for the two- and three-particle 
continua resulting from the approximate one-triplon dispersion $\omega_1(K)$ in
the case of the generator  $F_{\textrm{gs}}(l)$ for $x=0.5$ and
$y=0.1$. The three-particle continuum can be determined in 
analogy to the two-particle continuum. One only has to
replace one one-triplon dispersion on the right hand side of
\eref{eq:2p_cont} by energies of the two-particle continuum. 

\begin{figure}
  \centering
  \includegraphics[width=0.65\columnwidth]{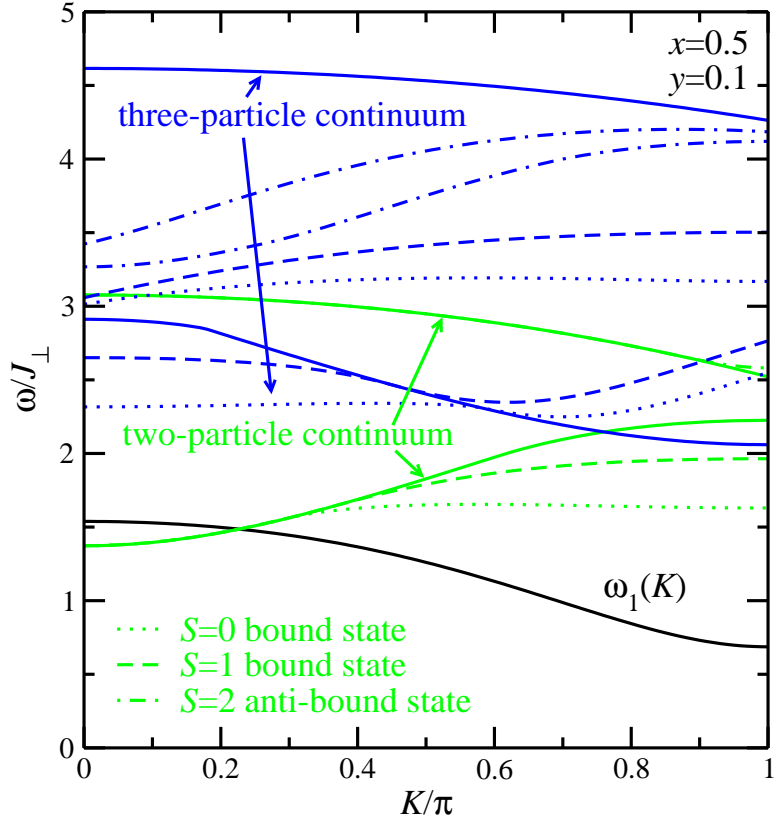}
  \caption{\label{fig:3bound}
    Two- and three-triplon continua of the asymmetric spin-${1}/{2}$
    Heisenberg ladder with $x=0.5$ and $y=0.1$. The solid green lines 
    represent the lower and upper boundaries of the two-particle continuum. 
    The other green lines represent two two-particle bound states and one 
    two-particle anti-bound state. The blue lines illustrate the boundaries of 
    the three-particle continuum, where the type of the lines correspond to 
    the two-particle state(s) used to determine the three-particle continuum.}
\end{figure}

The boundaries of these continua are shown in \fref{fig:3bound} by solid lines.
Additionally, this figure shows the boundaries of the
three-particle continua\footnote{The name three-particle
  continuum refers to the fact that the corresponding states consist
  of three triplons in the basis of the effective Hamiltonian.}
emerging from the combination of the approximate one-triplon
dispersion  $\omega_1(K)$  and a certain approximate two-particle
(anti-)bound state. Here we only use the region where the
corresponding (anti-)bound
state is well-defined, i.e., does not merge with the continuum.

 The two-particle bound states and the two-particle anti-bound state shown in \fref{fig:3bound} are calculated by diagonalizing the
effective Hamiltonian $H_{\textrm{eff}}$ in the subspace spanned by the
single-triplon states 
\numparts
  \begin{equation}
    \label{eq:2p_sub_a}    
    \ket{K,\alpha} := \frac{1}{\sqrt{\mathcal{N}}} \sum_r \e^{\rmi K r}
    \ket{r,\alpha} 
  \end{equation}
  and the two-triplon states
  \begin{equation}
    \label{eq:2p_sub_b}
    \ket{K,\alpha}\ket{d,\beta} := \frac{1}{\sqrt{\mathcal{N}}} \sum_r \e^{\rmi
      K\left(r + \frac{d}{2} \right)} \ket{r,\alpha} \ket{r + d,\beta} 
  \end{equation}
\endnumparts
with $0<d < 120$ for each given value of $K$ 
(for details see \sref{app:analysis})\footnote{Here we considered $d<120$ 
only to be consistent with  the later calculations which also include the 
three-particle space.}. 

Since the subspace spanned by the states
\eref{eq:2p_sub_a} and \eref{eq:2p_sub_b} is not separated from higher triplon states
(cf.\ \fref{fig:subspaces}b we obtain -- as for the one-triplon
dispersion  $\omega_1(K)$ -- only an approximation for the
bound states. Consequently, the depicted continua only represent 
approximations as well. The restriction of the relative distance $d$
in the two-triplon subspace \eref{eq:2p_sub_b} is less
important. Increasing $d$ does not change the results perceivably. 

As in the symmetric case $x=0.5$ and 
$y=0$ \cite{uhrig96b,uhrig96be,damle98,SK98,trebs00,KSGU01}, 
two two-particle bound 
states -- one with total spin $S=0$ and one with total spin
$S=1$ -- and an anti-bonding state with total spin $S=2$ exist. The boundaries
of the continua shown in \fref{fig:3bound} help to understand the shape of 
the spectral densities calculated in the following section.

Note that the whole complex structure of the low energy spectrum shown
in \fref{fig:3bound} follows from the one-triplon dispersion,
the triplon-triplon interaction,
and from the  diagonalization of the effective Hamiltonian within the 
subspace \eref{eq:2p_sub_a}-\eref{eq:2p_sub_b}.

\subsection{Spectral density}
\label{sec:spectral_density}
In this subsection we improve the results presented in the former section for 
the one-triplon
dispersion which we obtained by the generator $F_{\textrm{gs}}(l)$ for $x=0.5$
and $y=0.1$. This is achieved by including interactions with three-triplon 
states. To describe  triplon decay we calculate the zero temperature spectral 
density.

We start by analyzing the frequency and momentum resolved
retarded zero temperature Green function
\begin{equation}
  G\left(K,\omega\right) = \lim_{\delta \rightarrow 0^+}\bra{K,z}
  \big[{\omega-\left(H_{\textrm{eff}}-E_0\right)+ \rmi \delta}\big]^{-1} 
\ket{K,z}.
\end{equation}
The spectral density $S(K,\omega)$ follows by taking the negative imaginary 
part of $G\left(K,\omega\right)$ divided by $\pi$
\begin{equation}
  \label{eq:spectral_density}
  S\left(K,\omega\right) = -\frac{1}{\pi} \Im \left[ G\left(K,\omega\right)\right].
\end{equation}
The Green function is evaluated by tridiagonalization (Lanczos
algorithm) which leads
to the continued fraction representation \cite{Z61,M65,GB87,PW85,VM94}
\begin{equation}
  \label{eq:green_frac}
  G(K,\omega) = \frac{1}{\omega - a_0(K) - \frac{b_1(K)^2}{\omega - a_1(K) - 
\frac{b_2(K)^2}{\ldots}}}. 
\end{equation}
The coefficients $a_i(K)$ and $b_i(K)$ are calculated by repeated application
of $H_{\textrm{eff}}-E_0$ on the initial state $\ket{K,z}$
with wave vector $K$, spin $S=1$, and $S_z$ component $m=0$  
(for details see \sref{app:lanczos}). Note that the continued fraction in the denominator on the
right hand side of \eref{eq:green_frac} (proportional to $b_1(K)^2$)
can be taken as a standard self-energy whose imaginary part
determines the decay rate. In this respect our approach is
not so different from the one in \cite{BFS98}.

In all practical calculations we have to restrict ourselves to a certain
subspace. For this calculations we considered the subspace spanned by
\numparts
  \begin{eqnarray}
    \label{eq:3_sub_a}  
    \fl \ket{K,\alpha} := \frac{1}{\sqrt{\mathcal{N}}} 
    \sum_r \e^{\rmi K r}
    \ket{r,\alpha}, \\ \fl
    \label{eq:3_sub_b}  \ket{K,\alpha}\ket{d,\beta} :=
    \frac{1}{\sqrt{\mathcal{N}}} \sum_r \e^{\rmi
      K\left(r + \frac{d}{2} \right)} \ket{r,\alpha} \ket{r + d,\beta} 
  \end{eqnarray}
  and
  \begin{equation}
    \eqalign{
      \label{eq:3_sub_c} \fl  \ket{K,\alpha}\ket{d_1,\beta}\ket{d_2,\gamma} :=& 
      \frac{1}{\sqrt{N}}
      \sum_r \e^{\rmi K \left(r + \frac{2 d_1 + d_2}{3} \right) }\ket{r,\alpha} \ket{r + d_1,\beta} \ket{r + d_1 + d_2,\gamma}
    }
  \end{equation}
\endnumparts
with $d,d_1,d_2>0$ and $d,d_1+d_2 < 120$. Note that \eref{eq:3_sub_a}
is the one-triplon state for fixed $K$ and $\alpha$,
\eref{eq:3_sub_b} the two-triplon scattering states, and
\eref{eq:3_sub_c} the three-triplon scattering states. 
Thus we only need the restricted effectiveHamiltonian 
\begin{eqnarray}
  \eqalign{
    \label{H_res}
    \fl H_{\textrm{eff}}^{\textrm{res}} = J_{\perp}\left( H_1^1(\infty) +
    H_2^2(\infty) +H_3^3(\infty) + H_1^2(\infty) + H_2^1(\infty) + H_1^3(\infty) + H_3^1(\infty) 
    \right.
    \\
    +\left.H_2^3(\infty) + H_3^2(\infty) \right)
  }
\end{eqnarray}
The action of this restricted effective Hamiltonian 
$H_{\textrm{eff}}^{\textrm{res}}$ on the subspace \eref{eq:3_sub_a}-\eref{eq:3_sub_c} is presented in \sref{app:analysis}. More details about the calculation of the spectral 
density are given in \sref{app:lanczos}.

\begin{figure*}[ht]
  \begin{minipage}{0.645\textwidth}
    \includegraphics[width=1\textwidth]{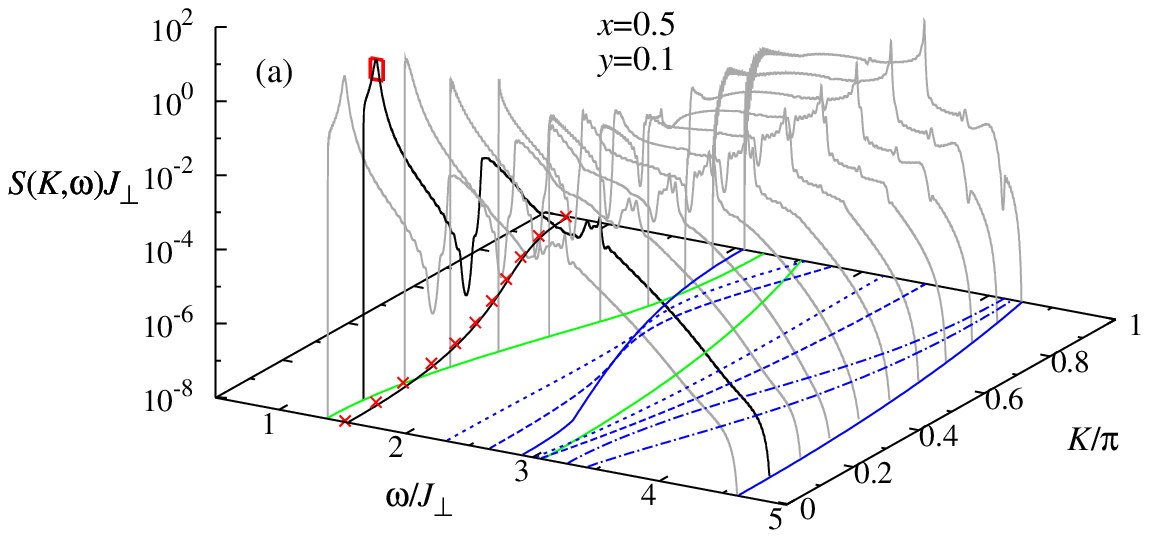}
  \end{minipage}
  \begin{minipage}{0.345\textwidth}
    \includegraphics[width=0.85\textwidth]{figure10b}
  \end{minipage}\\  
  \vspace{5mm}
  \begin{minipage}{0.495\textwidth}
    \includegraphics[width=0.95\textwidth]{figure10c}
  \end{minipage}
  \begin{minipage}{0.495\textwidth}
    \includegraphics[width=0.95\textwidth]{figure10d}
  \end{minipage}  
  \caption{\label{fig:spectral}
    Spectral properties of the asymmetric
    spin-${1}/{2}$ ladder. Panels (a), (b) and (c) show the
    spectral density of the asymmetric
    spin-${1}/{2}$ ladder for $x=0.5$ and $y=0.1$. In Panel (b)
    a Lorentzian is fitted to the spectral density. The described
    quasiparticle exhibits a inverse lifetime of $\Gamma \approx
    0.0402 J_{\perp}$. Panel (d) compares the results of
    \fref{fig:disp}b with the red crosses depicting the
    renormalized one-triplon dispersion obtained by using the generator
    $F_{\textrm{gs}}(l)$ and (tri)diagonalization in
    the subspace \eref{eq:3_sub_a}-\eref{eq:3_sub_c}. This renormalized one-triplon
    dispersion is also depicted in Panel (a).}
\end{figure*}

\Fref{fig:spectral}a shows the spectral density $S(K,\omega)$ for
$x=0.5$ and $y=0.1$. 
To keep track of this spectral density we display
the results presented in \fref{fig:3bound} also in the
$\omega,K$-plane of \fref{fig:spectral}a. For $K=0$
and $K={\pi}/{10}$ decaying triplons are observed. Their density lies 
in the vicinity of the approximate one-triplon dispersion. The region framed 
in  red is shown in detail in \fref{fig:spectral}b. In this region we can 
fit our data by a Lorentzian 
\numparts
  \begin{equation}
    L(\omega)=\frac{A}{\pi}\frac{\left({\Gamma}/{2}\right)^2}
    {\left(\omega  - \omega_{0} \right)^2 + {\left({\Gamma}/{2}
	\right)^2}}
  \end{equation}
  with
  \begin{eqnarray}
    A \approx& 1.0848/J_\perp \\
    \omega_0 \approx&  1.4938 J_{\perp} 
  \end{eqnarray}
  and the inverse lifetime
  \begin{equation}
    \Gamma \approx 0.0402  J_{\perp}.
  \end{equation}
\endnumparts

For clarity, \fref{fig:spectral}c shows the spectral density at
$K={\pi}/{10}$ on logarithmic scale. Besides the strong one-triplon 
peak at $\omega_0 \approx  1.4938 J_{\perp}$ the spectral density increases
distinctly at the beginning of the three-particle continuum involving
the $S=0$ bound state. Another rise of the spectral density
occurs at the upper end of the three-particle continuum involving
the $S=1$ bound state. At the beginning of the three-particle continuum 
involving the $S=2$ anti-bound state the spectral density drops notably. 
This illustrates that the existence of bound and anti-bound states
influences the form of the spectral density significantly.
Note that an additional channel, for instance contributions
from scattering states of a single triplon and a triplon-triplon bound state,
does not always imply an increase of the spectral density. It
may also lead to a significant decrease. We attribute this phenomenon
to destructive interference. That means the additional channel
interferes destructively with the already existing channel so that
a net decrease is engendered.

Finally, we want to discuss the shift of the one-triplon
dispersion caused by the hybridization of two- and three-particle
states (cf.\ \eref{eq:3_sub_a}-\eref{eq:3_sub_c}). In the sequel we call this shifted 
one-triplon dispersion as the renormalized one-triplon dispersion. 
The results shown in \fref{fig:spectral}d 
are partly those of \fref{fig:disp}b. In addition,
the red crosses depict results obtained from a tridiagonalization
after the CUT induced by $F_{\textrm{gs}}(l)$.

In the region of the Brillouin zone where the quasiparticles are
well-defined we obtain the renormalized one-triplon dispersion by
fixing the total momentum $K$ and calculating the lowest eigenvalue of
the effective Hamiltonian $H_{\textrm{eff}}$ in the subspace
\eref{eq:3_sub_a}-\eref{eq:3_sub_c}. The results are represented in \fref{fig:spectral}a and \fref{fig:spectral}d by red crosses.   
To good accuracy, we retrieve the results obtained before by
$F_{\textrm{pc}}(l)$ and $F_{\textrm{gs,1p}}(l)$ in the region
without decay. Therefore, it is sufficient to consider the
subspace \eref{eq:3_sub_a}-\eref{eq:3_sub_c} if one wants to describe the one-triplon 
dispersion of the
asymmetric spin-${1}/{2}$ Heisenberg ladder with $x=0.5$ and
$y=0.1$ by using the generator $F_{\textrm{gs}}(l)$. 

Note that the calculation in the subspace \eref{eq:3_sub_a}-\eref{eq:3_sub_c} does not lead
to the correct band edges of the triplon continuum because the shift of the
one-triplon dispersion makes itself felt only if we 
included four-particle states as well. Hence this kind of calculation
is not fully self-consistent. There are possibilities to
achieve consistency between the one-triplon dispersion and the
band edges of the continua. But this issue 
is beyond the scope of the present article.

In the region where the one-triplon dispersion hybridizes with the
two-triplon continuum the red crosses shown in \fref{fig:spectral}a and \fref{fig:spectral}d indicate the
energy with the maximum spectral intensity $S(K,\omega)$.
These energies represent what is usually seen as the energy of
a quasiparticle with finite life-time. The energies determined in this
way lie between what is obtained from $F_{\textrm{gs}}(l)$
in the one-triplon sector (blue line in \fref{fig:spectral}d)
and what is obtained from $F_{\textrm{pc}}(l)$ or from 
$F_{\textrm{gs,1p}}(l)$.

We emphasize, that the advantage of the generator $F_{\textrm{gs}}(l)$ 
compared to the
generator  $F_{\textrm{gs,1p}}(l)$ or $F_{\textrm{pc}}(l)$ is that
also the quasiparticle decay is described in the region of the
Brillouin zone where the one-triplon dispersion merges with the
two-triplon continuum. The generator $F_{\textrm{gs}}(l)$
avoids rearrangement processes during the flow which lead to a potentially
misleading quasiparticle picture. Thereby the CUT becomes more robust.
Hence the proposed adapted generator achieves the goal from the
outset to describe decaying quasiparticles properly.

\section{Summary}
\label{sec:summary}
In the present paper, we introduced an approach based on continuous
unitary transformations to describe systems
with unstable quasiparticles. The main idea is to use a generator
formulated in second quantization
which leads to an effective model where only the ground state is
isolated from the rest of the Hilbert space. 

In the first part of this paper we described the properties of this
adapted generator and discussed similarities and differences with other
generators. Additionally, we derived the adapted generator in the context
of variational calculations. All considerations were completely
general and did neither depend on the model under study nor on the actual
realization of the continuous unitary transformation. 
Thus we expect that generally an analogous modification of the unitary
transformation can also be used in other approaches, for instance in 
high-order series expansions by orthogonal transformations \cite{oitmaa06},
 to capture resonant behavior. Not the life time will be 
accessible directly by a series, but the effective Hamiltonian.
Then a subsequent analysis by variational methods (as presented in
the present work) or by diagrammatic approaches has to be used.

In the second part of this paper we illustrated the theoretical
deliberations for the asymmetric antiferromagnetic
spin-${1}/{2}$ Heisenberg ladder. This model shows  spontaneous
triplon decay into two-triplon scattering states. The strength of
this decay is controlled by the frustrating diagonal interaction 
$J_\textrm{diag}$. A more comprehensive study of this
particular model is left to ongoing research.

 We used the generator $F_{\textrm{gs}}(l)$ which only
isolates the ground state and an additional
Lanczos tridiagonalization in a variational 
subspace which consisted of states containing
up to three triplons. We showed that
in this way the resonance behavior of the decaying
triplon can be described explicitly. 
The continuous unitary transformation was realized
in a self-similar way. 

In conclusion, we extended the range of applications of continuous
unitary transformation to systems which exhibit unstable quasiparticles.
We expect that an analogous extension can also be implemented
for other unitary transformations.

\ack
\label{sec:acknowledgements}

We want to thank K. P. Schmidt, N. Drescher and C. Raas for many
fruitful discussions. This work was supported by the NRW
Forschungsschule ``Forschung mit Synchrotronstrahlung in den Nano- und
Biowissenschaften''.

\appendix

\section{Ordering of the generator $F_{\textrm{pc}}(l)$}
\label{app:ordering}
The generator $F_{\textrm{pc}}(l)$ sorts the eigenvalues $E_i$ in 
ascending order of
the particle
number $q_i$ of the corresponding eigenvectors such that $q_i>q_j
\Rightarrow E_i > E_j$. Particularly, this implies that the vacuum
state
$\ket{0}$ of the effective Hamiltonian represents the ground state of
the system. In the following, we derive this statement.

In an eigenbasis of the operator $Q$, which counts the number of
(quasi)particles of a given state, the generator $F_{\textrm{pc}}(l)$ is 
given by
\begin{equation}
  \label{eq:eta_mku_matrix}
  \mathcal{F}_{i,j}(l) = \sgn \left( q_i - q_j \right) \mathcal{H}_{i,j}(l),
\end{equation}
where $q_i$ and $q_j$ are eigenvalues of the operator $Q$.
In general the eigenspace for a given number of (quasi)particles $q_i$ has a 
large dimension which is infinite for infinite system size, but for the 
purpose of the present derivation we stick to finite dimensional Hilbert 
spaces. We use the convention that $\mathcal{H}_{i,j}(l)$ stands
not only for a single matrix element but for the whole submatrix of the
Hamiltonian $\mathcal{H}$ which connects the eigenspace belonging to the
eigenvalue $q_i$ with the eigenspace belonging to the
eigenvalue $q_j$. Therefore $\mathcal{H}_{i,j}(l)$ is given by a matrix with
the dimension $D_i \times D_j$, where $D_i$ is the dimension of the eigenspace 
$Q=q_i$. 

Using the generator \eref{eq:eta_mku_matrix} the general flow equation
\eref{eq:flow_equation} yields the matrix equation
\begin{eqnarray}
  \label{eq:mku_flow}
  \eqalign{
  \fl \partial_l \mathcal{H}_{i,j}(l)
  =- \sgn \left( q_i - q_j \right) \left(
  \mathcal{H}_{i,i}(l) \mathcal{H}_{i,j}(l) - \mathcal{H}_{i,j}(l)
  \mathcal{H}_{j,j}(l) \right)\\ 
  + \sum\limits_{k \neq i,j} \left( \sgn \left( q_i - q_k \right)+\sgn\left( q_j
  - q_k \right) \right) \mathcal{H}_{i,k}(l) \mathcal{H}_{k,j}(l).}
\end{eqnarray}
Since the effective model will be block diagonal, all off-diagonal matrices
$\mathcal{H}_{i,j}(l)$ with $i \neq j$ have to vanish for $l \rightarrow
\infty$. Hence for large $l$ the equation \eref{eq:mku_flow} is dominated by the
first term on the right hand side where the off-diagonal matrices only appear
linearly. So for large $l$ the asymptotic behavior of 
\eref{eq:mku_flow} is given by 
\begin{equation}
  \label{eq:mku_flow_2}
    \partial_l \mathcal{H}_{i,j}(l) =
 -\sgn \left( q_i - q_j \right) \left(
    \mathcal{H}_{i,i} \mathcal{H}_{i,j}(l) - \mathcal{H}_{i,j}(l)
    \mathcal{H}_{j,j} \right) + {\Or} \left(\mathcal{H}_{i,j}^2(l) \right).
\end{equation}
Note that within this approximation $\partial_l \mathcal{H}_{i,i}(l) = 0$
$\forall i$, so that we can neglect the $l$-dependence of $\mathcal{H}_{i,i}(l)$ and $\mathcal{H}_{j,j}(l)$.
Without loss of generality we assume in the following that
$q_i > q_j$. Then \eref{eq:mku_flow_2} yields
\begin{equation}
  \label{eq:mku_flow_3}
  \partial_l \mathcal{H}_{i,j}(l) = -\left(
    \mathcal{H}_{i,i} \mathcal{H}_{i,j}(l) - \mathcal{H}_{i,j}(l)
    \mathcal{H}_{j,j} \right) + {\Or} \left(\mathcal{H}_{i,j}^2(l) \right).
\end{equation}
The matrix $\mathcal{H}_{i,i}$ and the matrix $\mathcal{H}_{j,j}$ are
Hermitian, thus unitary transformations $\mathcal{U}_i$ and
$\mathcal{U}_j$ exist which diagonalize $\mathcal{H}_{i,i}$ and
$\mathcal{H}_{j,j}$, respectively. We will denote this diagonal matrices by
$\mathcal{D}_i:=\mathcal{U}_i^{\dag}\mathcal{H}_{i,i}\mathcal{U}_i
$ and
$\mathcal{D}_j:=\mathcal{U}_j^{\dag}\mathcal{H}_{j,j}\mathcal{U}_j$. By
multiplying \eref{eq:mku_flow_3} from the left by
$\mathcal{U}_i^{\dag}$ and from the right by $\mathcal{U}_j$
one obtains
\begin{eqnarray}
  \label{eq:mku_flow_4}
  \partial_l
  \mathcal{\widetilde{H}}_{i,j}(l) = -\left(
    \mathcal{D}_{i} \mathcal{\widetilde{H}}_{i,j}(l) -
    \mathcal{\widetilde{H}}_{i,j}(l) \mathcal{D}_{j} \right) + {\Or} \left(\mathcal{\widetilde{H}}_{i,j}^2(l) \right),
\end{eqnarray}
where $\mathcal{\widetilde{H}}_{i,j}(l) :=
\mathcal{U}_i^{\dag}\mathcal{H}_{i,j}(l)\mathcal{U}_j $. According to
\eref{eq:mku_flow_4} the $(n,m)$ matrix element of 
$\mathcal{\widetilde{H}}_{i,j}(l)$ satisfies 
\begin{eqnarray}
  \label{eq:mku_flow_5}
  \eqalign{
    \partial_l\left(\mathcal{\widetilde{H}}_{i,j}(l)\right)_{n,m} &= -
    \sum\limits_{k} \Big(\mathcal{D}_{i}\Big)_{n,k}
    \left(\mathcal{\widetilde{H}}_{i,j}(l)\right)_{k,m}
    +   \sum\limits_{k}\left(\mathcal{\widetilde{H}}_{i,j}(l)\right)_{n,k}
    \left(\mathcal{D}_{j}\right)_{k,m}\\
    &= -\left( \Big(\mathcal{D}_{i}\Big)_{n,n} -
      \left(\mathcal{D}_{j}\right)_{m,m} \right) 
      \left(\mathcal{\widetilde{H}}_{i,j}(l)\right)_{n,m}
  }
\end{eqnarray}
in linear order in the non-diagonal matrices.
Since $\mathcal{H}_{i,j}(l)$ vanishes for $l \rightarrow \infty$,
$\mathcal{\widetilde{H}}_{i,j}(l)$ must vanish as well. Therefore, for large $l$ 
the inequality 
\begin{equation}
\label{eq:D-inequ}
   \Big(\mathcal{D}_{i}\Big)_{n,n} -
  \left(\mathcal{D}_{j}\right)_{m,m}  > 0
\end{equation} 
must be fulfilled for all $n,m$ for which the matrix elements
$(\mathcal{\widetilde{H}}_{i,j}(l))_{n,m}\nequiv 0$ are non-zero. 
This implies that all
eigenvalues of $\mathcal{D}_{i}$ must be larger than the eigenvalues of
$\mathcal{D}_{j}$. Thus, the eigenvalues are sorted in ascending order of the 
particle number of the corresponding eigenvectors, as asserted above. 

Note  that this ordering does not need to occur, if the corresponding 
eigenvectors  are not connected to each other by a finite matrix element of 
the Hamiltonian  for $l$ large, but not infinite. If 
$\mathcal{\widetilde{H}}_{i,j}(l)=0$
for all $l$ or for $l>l_0<\infty$ the argument to derive \eref{eq:D-inequ}
from \eref{eq:mku_flow_5} does not hold.
For example, this is the case when the system 
exhibits symmetries which prevent certain subspaces to be linked, as we see for the symmetric spin ladder.

\section{Transformation of subspaces}
\label{app:similarities}
\subsection{Ground state}
\label{app:sub:gs}
In \sref{subsec:properties} we argue that all generators
$F_{\textrm{pc}}(l)$, $F_{\textrm{gs}}(l)$, $F_{\textrm{mgs}}(l)$, and
$F_{\textrm{gs,1p}}(l)$) considered so far transform the vacuum state
$\ket{0(l)}$ in the same way if the flow equation is solved exactly. Here we 
prove this statement.

Previously, we defined the $l$-dependent Hamiltonian by $H(l):=U^{\dag}(l) H
U(l)$. Alternatively, we can keep the operators constant but make the states 
$l$-dependent.  This is in complete analogy to passing from the Heisenberg 
picture to the Schr\"odinger picture.
Hence, the $l$-dependence of the vacuum state is given by
$\ket{0(l)}=U(l)\ket{0}$ and the generator is given by $F(l)= -U^{\dag}(l)
\left( \partial_l U(l) \right)$. Thus, for the derivative of $\ket{0(l)}$ 
it follows 
\begin{eqnarray}
  \eqalign{
    \partial_l \ket{0(l)} &= \partial_l U(l) \ket{0} \\
    &= U(l) \underbrace{U^{\dag}(l) \left( \partial_l U(l)
      \right)}_{=-F(l)} \ket{0} \\
    &= - U(l)F(l) \ket{0}.
  }
\end{eqnarray}
Introducing a basis $\left\{ \ket{i} \right\}$ yields
\begin{eqnarray}
  \eqalign{
    \label{eq:0(l)}
    \partial_l \ket{0(l)} =& - \sum_{i} U(l) \ket{i} \underbrace{\bra{i} F(l) \ket{0}}_{=F_{i,0}(l)}.
  }
\end{eqnarray}
The key observation is that for all considered generators the matrix element $F_{i,0}(l)$ is the same, namely
\begin{eqnarray}
  \label{eq:def_g_i0}
  F_{i,0}(l) = 
  \cases{H_{i,0}(l) & for $i>0$\\ 0 & for $i=0$}.
\end{eqnarray}
Applying \eref{eq:def_g_i0} to \eref{eq:0(l)} yields
\begin{eqnarray}
  \eqalign{
    \partial_l \ket{0(l)} &= - \sum_{i \neq 0} U(l) \ket{i} \bra{i} H(l)
      \ket{0}  \\
    &= - \left(\sum_{i} U(l) \ket{i} \bra{i} H(l)  \ket{0} \right) + U(l) \ket{0} \bra{0} H(l) \ket{0}.
  }
\end{eqnarray}
Shifting the $l$-dependency to the vacuum state and using the equality
$U(l)U^{\dag}(l) \equiv 1 $ provides us with
\begin{eqnarray}
  \eqalign{
    \label{eq:final_0(l)}
    \partial_l \ket{0(l)} &=  -H \ket{0(l)} +  \ket{0(l)} \bra{0(l)} H
    \ket{0(l)}\\
    &=  \Big[ P_{0}(l), H \Big]  \ket{0(l)}
  }
\end{eqnarray}    
with the $l$-dependent projector $P_{0}(l)=\ket{0(l)} \bra{0(l)}$.
According to \eref{eq:final_0(l)} the derivative of $\ket{0(l)}$ only depends 
on $\ket{0(l)}$
itself and the initial Hamiltonian $H$. Therefore, the considered generators
 all transform the vacuum state $\ket{0(l)}$ in the same way. The essential 
point of the proof is that for all considered generators the matrix
elements $F_{i,0}(l)$ are defined identically by $\eref{eq:def_g_i0}$. 
Note, however, that the statement, that all generators treat $\ket{0}$ alike, 
does no longer hold if approximations (truncations) are introduced.

\subsection{One-particle space}
\label{app:sub:1p}
The proof presented in the previous subsection can be generalized. Since the
action of the generator $F_{\textrm{pc}}(l)$ and the generator
$F_{\textrm{gs,1p}}(l)$ is also the same on the one-particle subspace, one
can prove that they also transform all one-particle states in the same way. In the following we characterize the states by their number of
(quasi)particles, so it is useful to use an eigenbasis $\left\{\ket{i}\right\}$ of the (quasi)particle number operator $Q$. The number
of (quasi)particles of a state $\ket{i}$ is denoted by $q_i$.
Consider the derivative of all states with at most one particle
\begin{eqnarray}
  \eqalign{
    \partial_l \sum_{\makebox[0pt]{$\scriptstyle i: q_i \leq 1$}} \ket{i(l)}&= \sum_{i: q_i \leq 1} \partial_l U(l)\ket{i} \\
    &= - \sum_{\makebox[0pt]{$\scriptstyle i: q_i \leq 1$}} U(l)F(l) \ket{i}\\
    &= - \sum_{\makebox[0pt]{$\scriptstyle i: q_i \leq 1,j$}} U(l) \ket{j} \underbrace{\bra{j} F(l) \ket{i}}_{=F_{j,i}(l)}.
  }
\end{eqnarray}
For both generators the matrix elements $F_{j,i}(l)$ with $q_i \leq 1$ are
given by
\begin{eqnarray}
  \label{eq:def_g_ij}
  F_{j,i}(l) = \sgn \left( q_j - q_i \right) H_{j,i}(l)
\end{eqnarray}
according to \eref{eq:eta_pc} and \eref{eq:eta_gs_1p}.
Hence we have
\begin{eqnarray}
  \eqalign{
    \label{eq:1p(l)}
   \fl \partial_l \sum_{i: q_i \leq 1} \ket{i(l)}= -
    \sum_{\makebox[0pt]{$\scriptstyle i: q_i \leq 1, j: q_j > 1$}}
    U(l) \ket{j} {\bra{j} H(l) \ket{i}}\\
    - \sum_{\makebox[0pt]{$\scriptstyle i: q_i \leq 1, j: q_j \leq 1$}}\sgn\left( q_j - q_i \right)
    U(l) \ket{j} {\bra{j} H(l) \ket{i}}
  }
\end{eqnarray}
To the first part on the right hand side of \eref{eq:1p(l)}
we add all missing contributions with $ q_j \leq 1$. Hence we arrive at
\begin{eqnarray}
  \eqalign{
    \fl \partial_l \sum_{i: q_i \leq 1} \ket{i(l)}= -
    \sum_{\makebox[0pt]{$\scriptstyle i: q_i \leq 1$}} U(l) H(l)
    \ket{i}
    +\sum_{\makebox[0pt]{$\scriptstyle i: q_i \leq 1, j: q_j \leq 1$}}
    U(l) \ket{j} {\bra{j} H(l) \ket{i}}\\
    - \sum_{\makebox[0pt]{$\scriptstyle i: q_i \leq 1, j: q_j \leq 1$}}\sgn\left( q_j - q_i \right)
    U(l) \ket{j} {\bra{j} H(l) \ket{i}}.
  }
\end{eqnarray}
Just as in the previous subsection we shift the $l$-dependence from the
Hamiltonian $H(l)$ to the states
\begin{eqnarray}
  \eqalign{
    \fl \partial_l \sum_{\makebox[0pt]{$\scriptstyle i: q_i \leq 1$}} \ket{i(l)}= - \sum_{i: q_i \leq 1}  H \ket{i(l)}
    +\sum_{\makebox[0pt]{$\scriptstyle i: q_i \leq 1, j: q_j \leq 1$}}
    \ket{j(l)} {\bra{j(l)} H \ket{i(l)}}\\
    - \sum_{\makebox[0pt]{$\scriptstyle i: q_i \leq 1, j: q_j \leq 1$}}\sgn\left( q_j - q_i \right)
    \ket{j(l)} {\bra{j(l)}H \ket{i(l)}}.
  }
\end{eqnarray}
It follows that the transformation of the subspace
$\left\{\ket{i}\right\}$ with $q_i \leq 1$ is independent from all other states
$\left\{\ket{i}\right\}$ with $q_i > 1$. The transformation only depends on the
initial Hamiltonian $H$. Therefore, the generator $F_{\textrm{pc}}(l)$ and the generator
$F_{\textrm{gs,1p}}(l)$ transform the one-particle subspace in the same
way. Note that this proof is not restricted to the case $q_i \leq 1$ and
can easily be adapted to the case $q_i \leq n\in \mathbb{N}$.
The choice \eref{eq:eta_gs_1p} or \eref{eq:def_g_ij} has to be adapted accordingly, i.e., we have pass from 
$F_{\textrm{gs,1p}}(l)$ to $F_{\textrm{gs,$n$p}}(l)$ with
\begin{equation}
   \label{eq:eta_gs_np}
   F_{\textrm{gs,$n$p}}(l)=\sum_{p=0}^n\sum_{i>p}^N \left(H_{p}^{i}(l) -
     H_{i}^{p}(l)\right).
\end{equation}

\section{Lanczos tridiagonalization}
\label{app:lanczos}
To determine the spectral density \eref{eq:spectral_density} one has to calculate the coefficients $a_i(K)$
and $b_i(K)$ of the continued fraction representation of the Green
function \eref{eq:green_frac}. This can be done by the Lanczos
recursion scheme, for which we restrict our calculations to the subspace \eref{eq:3_sub_a}-\eref{eq:3_sub_c}.
Within this subspace
the recursion (Lanczos tridiagonalization)
\numpartsapp
\begin{eqnarray}
  \ket{\psi_0} =&\ket{K,z} \\
  \ket{\psi_1} =& \left({H_{\textrm{eff}}^{\textrm{res}} - a_0(K)}\right) \ket{\psi_0} \\
  \ket{\psi_2} =& \left({H_{\textrm{eff}}^{\textrm{res}} - a_1(K)}\right) \ket{\psi_1} -b_1(K)^2 \ket{\psi_0} \\
  \ket{\psi_3} =& \left({H_{\textrm{eff}}^{\textrm{res}} -
    a_2(K)}\right) \ket{\psi_2} -b_2(K)^2 \ket{\psi_1}
\end{eqnarray}
\begin{equation*}
  \vdots
\end{equation*}
with
\begin{eqnarray}
  a_i(K) &= \frac{\bra{\psi_i} H_{\textrm{eff}}^{\textrm{res}} \ket{\psi_i}}{ \braket{\psi_i | \psi_i}} \qquad&
  \textrm{for }  i=0,1,2,\ldots\\
  b_i(K)^2 &= \frac{\braket{\psi_i | \psi_i}}{ \braket{\psi_{i-1} | \psi_{i-1}}} \qquad&
  \textrm{for } i=1,2,3,\ldots\\
  b_0(K) &= 0
  \end{eqnarray}  
\endnumpartsapp
generates a set of orthogonal states $\ket{\psi_i}$. The action of the
restricted Hamiltonian \eref{H_res} in
the three-particles subspace \eref{eq:3_sub_a}-\eref{eq:3_sub_c} is given in \sref{app:analysis}. In the generated basis
$\left\{ \ket{\psi_i}\right\}$ the matrix of the restricted effective
Hamiltonian $H_{\textrm{eff}}^{\textrm{res}}$ is tridiagonal, where the $a_i(K)$ are
the diagonal matrix elements and the $b_i(K)$ are the elements on the second
diagonal. All other matrix elements are zero.

\begin{figure}
  \centering
  \includegraphics[width=0.65\columnwidth]{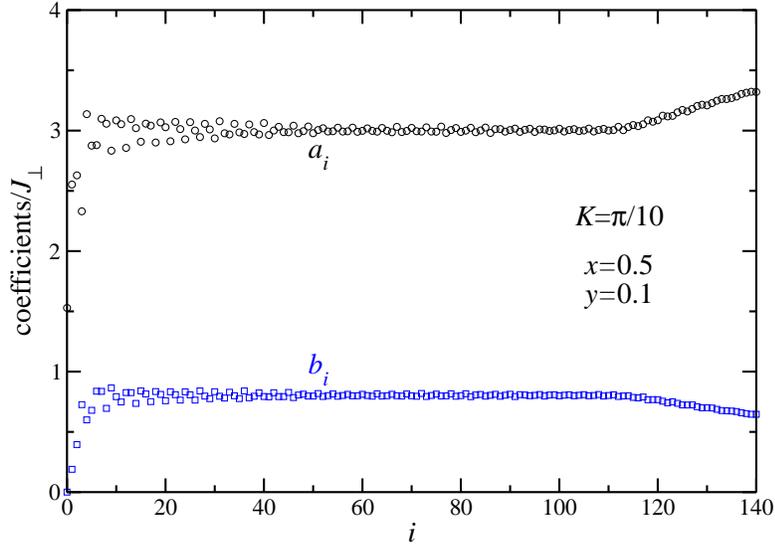}
  \caption{\label{fig:ab}
    Continued fraction coefficients  $a_i(K)$ and $b_i(K)$ for $x=0.5$, 
    $y=0.1$ and momentum $K={\pi}/{10}$. The restriction of the 
    considered subspace becomes conspicuous at $i\approx 115$.} 
\end{figure}

\Fref{fig:ab} shows the results for the coefficients $a_i(K)$ and
$b_i(K)$ for $x=0.5$, $y=0.1$ and $K={\pi}/{10}$.
First it appears that both coefficients $a_i(K)$ and
$b_i(K)$ ($i \lessapprox 115$) converge to fixed values
$a_{\infty}(K)$ and $b_{\infty}(K)$ as it should be for a bounded and
gapless spectral density of an infinitely large system \cite{PW85}. But for 
$i \gtrapprox 115$ both
coefficients start to change their values again noticeably. This is a
consequence of the fact that we had to restrict the relative distances
$d$ and $d_1+d_2$ to a maximum of $119$ rungs
(cf.\ \eref{eq:3_sub_a}-\eref{eq:3_sub_c}) in our numerical
calculations. Therefore, we only use the first 100 coefficients and terminate 
the continued fraction at $i=100$ as described in the following subsection.

\subsection{Termination}
\label{app:termination}
The spectral density $S(K,\omega)$ at fixed $K$ as obtained by a finite
continued fraction of the Green function \eref{eq:green_frac} has
poles at the zeros of the denominator. Thus, the spectral density
$S(K,\omega)$ is a collection of $\delta$-peaks. One standard approach to
obtain a continuous density is to introduce a slight broadening of
$S(K,\omega)$ via $\omega \rightarrow \omega + \rmi \delta$ with a
small real number $\delta$. This procedure corresponds to smearing out
$\delta$-peaks as Lorentzian functions of width $\delta$. The caveat is
that also all truly sharp features such as band edges or van Hove
sigularities are smeared out.
However, a notably improved resolution of $S(K,\omega)$ can be achieved by
introducing an appropriate termination of the continued fraction.

If we want to evaluate the continued fraction for the infinite large
system, we have to stop our recursion before the finiteness of the
considered subspace \eref{eq:3_sub_a}-\eref{eq:3_sub_c} becomes
conspicuous. Therefore, we compute the average value of $a_{i}(K)$ and
$b_{i}(K)$ for $i=80\ldots100$ (cf.\ \fref{fig:ab}) to obtain a good 
approximation for the limits
$a_{\infty}(K)$ and $b_{\infty}(K)$. From these limits the upper and
lower boundaries $E_{\textrm{u}}(K)$ and $E_{\textrm{l}}(K)$ of the
spectral density $S(K,\omega)$ are determined
via \cite{PW85}
\numpartsapp
  \begin{eqnarray}  
    E_{\textrm{u}}(K)=&  a_{\infty}(K) + 2b_{\infty}(K) \\
    E_{\textrm{l}}(K)=&  a_{\infty}(K) - 2b_{\infty}(K) \ .
  \end{eqnarray}
\endnumpartsapp
The existence of an upper boundary of the spectral density is a
consequence of our restriction to a subspace which contains
three quasiparticles at maximum.   
Finally, we use the directly calculated coefficients $a_{i}(K)$ and
$b_{i+1}(K)$ for $i=0\ldots99$ and subsequently the square root
terminator defined by the approximate limits $a_{\infty}(K)$ and
$b_{\infty}(K)$. By doing this, we assume that all following coefficients 
$a_{i}(K)$ and
$b_{i+1}(K)$ with  $i\geq100$ are constant. The square root terminator $T(K)$ 
is given by
\numpartsapp
  \begin{eqnarray}
    T&= \frac{1}{2b_{\infty}^2}
    \left(\omega-a_{\infty}-\sqrt{-D}\right) \qquad & \textrm{for } \omega \geq E_{\textrm{u}} \\
    T&= \frac{1}{2b_{\infty}^2} \left(\omega-a_{\infty}-\rmi\sqrt{D}\right)
      \qquad &\textrm{for } E_{\textrm{l}} \leq \omega \leq E_{\textrm{u}} \\
    T&= \frac{1}{2b_{\infty}^2}
    \left(\omega-a_{\infty}+\sqrt{-D}\right) \qquad &\textrm{for }
       \omega \leq E_{\textrm{l}}
  \end{eqnarray}
  with
  \begin{equation}
    D= 4 b_{\infty}^2 - \left(\omega-a_{\infty} \right)^2 \ ,
  \end{equation}
\endnumpartsapp
where we suppressed all $K$ dependences for the sake of simplicity. The last
considered coefficient $b_{100}(K)$ in \eref{eq:green_frac} is
multiplied by the appropriate terminator $T(K)$. The imaginary
part of the resulting expression yields the continuous part of the
spectral density $S(K,\omega)$. In this way we
can reliably approximate the thermodynamic limit of the spectral density 
$S(K,\omega)$
by calculations in a finite subspace.

\section{Analysis of the effective model}
\label{app:analysis}
Here we present details of the analysis of the
effective Hamiltonians for the asymmetric ladder generated by
CUTs. More details are given
in \cite{Knetter2003,KSU04twoleg,Kirschner2004} for the special case of a
particle conserving effective Hamiltonian.

The generators $F_{\textrm{pc}}(l)$ and $F_{\textrm{gs,1p}}(l)$ 
isolate the one-particle subspace from all other
subspaces (cf.\ \fref{fig:comparison}a and \fref{fig:comparison}c). 
Therefore,  the one-particle eigenvalues can be calculated
without considering states with a higher particle number. 

\begin{figure*}
  \includegraphics[width=1.0\textwidth]{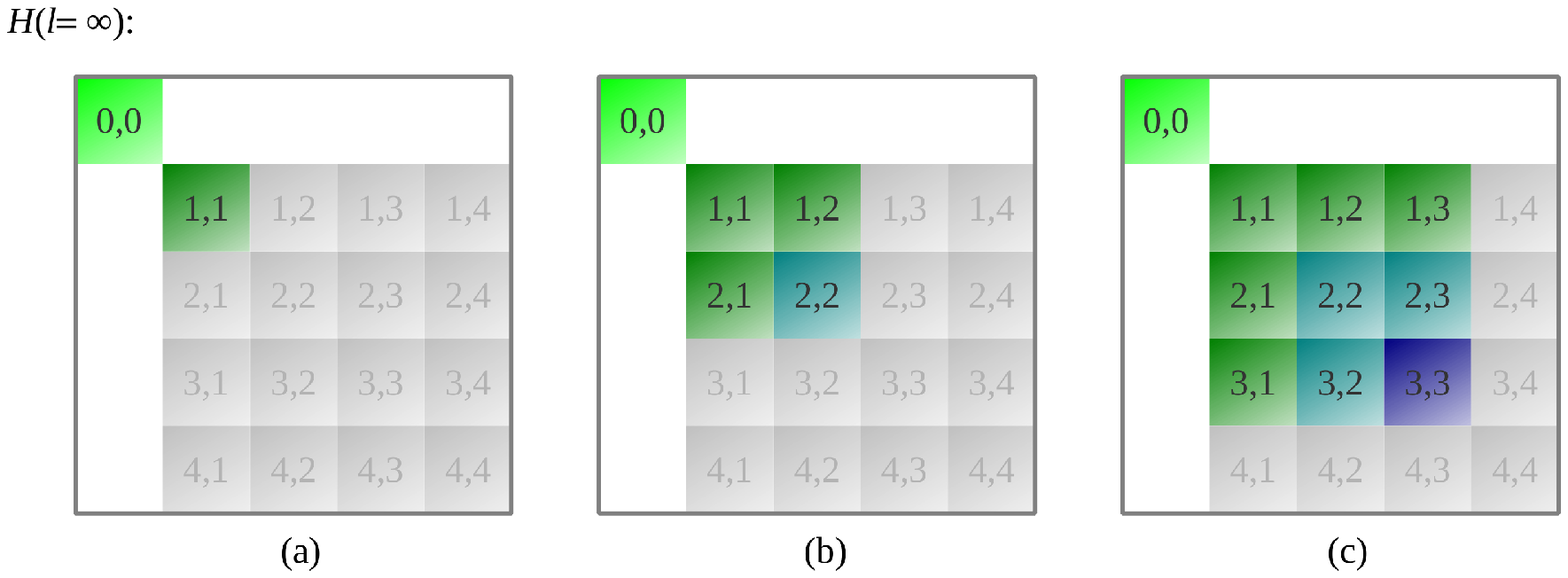}
  \caption{(Color online) Considered subspaces to analyze the effective 
    Hamiltonian obtained by the generator $F_{\textrm{gs}}(l)$. The generator
    $F_{\textrm{gs}}(l)$ only isolates the part $H_{0}^{0}(l)$. Colored
    blocks illustrate interactions which are included. Grey blocks illustrate
    neglected interactions. Panel (a) describes an analysis within the
    one-triplon subspace. Panel (b) describes an analysis within the
    one- and two-triplon subspace. Panel (c) describes an analysis within the
    one-, two- and three-triplon subspace, see \eref{eq:3_sub_a}-\eref{eq:3_sub_c}} 
  \label{fig:subspaces}  
\end{figure*}

The effective Hamiltonians obtained by the generator $F_{\textrm{gs}}(l)$
still contain interactions between the one-particle subspace and other
subspaces (cf.\ \fref{fig:comparison}b). Consequently, a
diagonalization in the one-particle subspace only gives an
approximation for the eigenvalues of the effective Hamiltonian,
namely an upper bound for the eigenvalues if the ground state energy
is sufficiently well described. The results for the eigenvalues 
can be improved by considering higher particle subspaces
as well (see \fref{fig:subspaces}).
In this paper we consider subspaces which consist of states which contain up 
to three triplons.

The Fourier transformed one-, two- and three-particle states
are given by \eref{eq:3_sub_a}-\eref{eq:3_sub_c}. In any practical calculation
the relative distances must be truncated to make the
subspace finite. Note that due to the
hard-core algebra it is not possible that two particles occupy the
same rung.
Below the action of the various parts of the effective
Hamiltonians are given. 

\subsection{$H_{1}^{1}$}
The action of the operator $H_{1}^{1}$ on the one-triplon state 
$\ket{K,\alpha}$ is given by
\begin{eqnarray}
  \label{eq:h_11|1}
  H_{1}^{1} \ket{K,\alpha}
  =& \sum_{{r},\alpha'} \e^{\rmi K {r}}
  c_{11;{r}}^{\alpha',\alpha} \ket{K,\alpha'}
\end{eqnarray}
with 
\begin{eqnarray}
   c_{11;{r}}^{\alpha',\alpha}:=& \bra{R,\alpha'} H_{1}^{1}
   \ket{R+r,\alpha}
\end{eqnarray}
and $R \in \mathbb{Z}$.
Note that due to the translational symmetry only the relative
distance $r$ between the states $\ket{R,\alpha'}$ and
$\ket{R+r,\alpha}$ occurs in the coefficient
$c_{11;{r}}^{\alpha',\alpha}$. In the special case of a SU(2) symmetric Hamiltonian the coefficients
$c_{11;{r}}^{\alpha',\alpha}$ obey the relation
$c_{11;{r}}^{\alpha',\alpha}= \delta_{\alpha',\alpha} c_{11;r}$.
The used truncation scheme
(cf. \sref{subsec:truncation}) causes
$c_{11;{r}}^{\alpha',\alpha}=0$ for $|r|>d_2$. All other coefficients
$c$ which appear in the following are affected by the truncation
scheme in an analogous way.

The action of the operator $H_{1}^{1}$ on the two-triplon state $\ket{K,\alpha}\ket{d,\beta}$
is given by
\numpartsapp
\begin{eqnarray}
  \fl  H_{1}^{1} \ket{K,\alpha} \ket{d,\beta} = \sum_{\makebox[0pt]{$\scriptstyle {r}>-d \atop \alpha'$}}
  c_{11;{r}}^{\alpha',\alpha}  \e^{\rmi
    K \frac{{r}}{2}} \ket{K, \alpha'} \ket{{r} +
    d, \beta}  \\
  +\sum_{\makebox[0pt]{$\scriptstyle {r}<-d \atop \alpha'$}} c_{11;{r}}^{\alpha',\alpha} \e^{\rmi
    K \frac{{r}}{2}} \ket{K, \beta} \ket{-\left( {r} + d \right), \alpha' }  \\
  + \sum_{\makebox[0pt]{$\scriptstyle {r}<d \atop \beta'$}} c_{11;{r}}^{\beta',\beta} \e^{\rmi K \frac{{r}}{2}} 
  \ket{K, \alpha} \ket{-\left({r}-d \right), \beta'} \\
  + \sum_{\makebox[0pt]{$\scriptstyle {r}>d \atop \beta'$}} c_{11;{r}}^{\beta',\beta} \e^{\rmi K \frac{{r}}{2}}
  \ket{K, \beta'} \ket{ r-d, \alpha } 
\end{eqnarray}
\endnumpartsapp

The action of the operator $H_{1}^{1}$ on the three-triplon state $\ket{K,\alpha}\ket{d_1,\beta}\ket{d_2,\gamma}$ 
is given by
\numpartsapp
\begin{eqnarray}
\fl  H_{1}^{1} \ket{K,\alpha} \ket{d_1,\beta} \ket{d_2,\gamma} 
  = \sum_{\makebox[0pt]{$\scriptstyle {r} >-d_1 \atop \alpha'$}} c_{11;{r}}^{\alpha',\alpha} \e^{\rmi K \frac{{r}}{3}}
  \ket{K,\alpha'} \ket{{r}+d_1,\beta} \ket{d_2,\gamma}  \\
  + \sum_{\makebox[0pt]{$\scriptstyle -(d_1+d_2)<{r}<-d_1 \atop \alpha'$}} c_{11;{r}}^{\alpha',\alpha} \e^{\rmi K \frac{{r}}{3}}
  \ket{K,\beta} \ket{-\left({r}+d_1\right),\alpha'} \ket{{r}+d_1+d_2,\gamma}  \\
  + \sum_{\makebox[0pt]{$\scriptstyle {r}<-(d_1+d_2) \atop \alpha'$}} c_{11;{r}}^{\alpha',\alpha} \e^{\rmi K \frac{{r}}{3}}
  \ket{K,\beta} \ket{d_2,\gamma} \ket{-\left({r}+d_1+d_2\right),\alpha'}  \\
  + \sum_{\makebox[0pt]{$\scriptstyle -d_2<{r}<d_1 \atop \beta'$}} c_{11;{r}}^{\beta',\beta} \e^{\rmi K \frac{{r}}{3}}
  \ket{K,\alpha} \ket{-({r}-d_1),\beta'} \ket{{r}+d_2,\gamma}  \\
  + \sum_{\makebox[0pt]{$\scriptstyle {r}>d_1 \atop \beta'$}} c_{11;{r}}^{\beta',\beta} \e^{\rmi K \frac{{r}}{3}}
  \ket{K,\beta'} \ket{{r}-d_1,\alpha} \ket{d_1+d_2,\gamma}  \\
  + \sum_{\makebox[0pt]{$\scriptstyle {r}<-d_2 \atop \beta'$}} c_{11;{r}}^{\beta',\beta} \e^{\rmi K \frac{{r}}{3}}
  \ket{K,\alpha} \ket{d_1+d_2,\gamma} \ket{-\left({r}+d_2\right),\beta'}  \\
  + \sum_{\makebox[0pt]{$\scriptstyle {r}<d_2 \atop \gamma'$}} c_{11;{r}}^{\gamma',\gamma} \e^{\rmi K \frac{{r}}{3}}
  \ket{K,\alpha} \ket{d_1,\beta} \ket{-\left({r}-d_2\right),\gamma'}  \\
  + \sum_{\makebox[0pt]{$\scriptstyle d_2<{r}<d_1+d_2 \atop \gamma'$}} c_{11;{r}}^{\gamma',\gamma} \e^{\rmi K \frac{{r}}{3}}
  \ket{K,\alpha} \ket{-\left({r}-d_1-d_2\right),\gamma'} \ket{{r}-d_2,\beta}  \\
  + \sum_{\makebox[0pt]{$\scriptstyle {r}>d_1+d_2 \atop \gamma'$}} c_{11;{r}}^{\gamma',\gamma} \e^{\rmi K \frac{{r}}{3}}
  \ket{K,\gamma'} \ket{{r}-d_1-d_2,\alpha} \ket{d_1,\beta} 
\end{eqnarray}
\endnumpartsapp

\subsection{$H_{1}^{2}$}
The action of the operator $H_{1}^{2}$ on the one-triplon state $\ket{K,\alpha}$
is given by
\begin{eqnarray}
  H_{1}^{2} \ket{K,\alpha}
  =& \sum_{\makebox[0pt]{$\scriptstyle {r},d' \atop \alpha',\beta'$}} c_{21;{r},d'}^{\alpha',\beta',\alpha} \e^{\rmi K\left({r}-\frac{d'}{2}\right)}
  \ket{K,\alpha'}\ket{d',\beta'}
\end{eqnarray}
with
\begin{eqnarray}
  c_{21;r,d'}^{\alpha',\beta',\alpha} :=& \bra{R,\alpha'}
  \bra{R+d',\beta'} H_{1}^{2} \ket{{R+r},\alpha} .
\end{eqnarray} 

The action of the operator $H_{1}^{2}$ on the two-triplon state $\ket{K,\alpha}\ket{d,\beta}$
is given by
\numpartsapp
\begin{eqnarray}
  \fl H_{1}^{2} \ket{K,\alpha} \ket{d,\beta}  
  =\sum_{\makebox[0pt]{$\scriptstyle {r} >-d+d' \atop
      d',\alpha',\gamma'$}}  c_{21;{r},d'}^{\alpha',\gamma',\alpha} \e^{\rmi K \left(\frac{4{r} + d -2d'}{6} \right)}
  \ket{K,\alpha'} \ket{d',\gamma'} \ket{{r} + d - d',\beta} 
  \\
  + \sum_{\makebox[0pt]{$\scriptstyle -d<{r} <-d+d' \atop
      d',\alpha',\gamma'$}} c_{21;{r},d'}^{\alpha',\gamma',\alpha} \e^{\rmi K \left(\frac{4{r} + d -2d'}{6} \right)}
  \ket{K,\alpha'} \ket{r+d,\beta}  \ket{-\left({r}+d \right)+d',\gamma'}  \\
  + \sum_{\makebox[0pt]{$\scriptstyle {r} <-d \atop
      d',\alpha',\gamma'$}}   c_{21;{r},d'}^{\alpha',\gamma',\alpha} \e^{\rmi K \left(\frac{4{r} + d -2d'}{6} \right)}
  \ket{K,\beta} \ket{-\left({r}+d \right),\alpha'}
  \ket{d',\gamma'}  \\
  + \sum_{\makebox[0pt]{$\scriptstyle {r} <d \atop
      d',\beta',\gamma'$}}  c_{21;{r},d'}^{\beta',\gamma',\beta} \e^{\rmi K
    \left(\frac{4{r} - d -2d'}{6} \right)}
  \ket{K,\alpha} \ket{-{r}+d,\beta'} \ket{d',\gamma'}  \\
  + \sum_{\makebox[0pt]{$\scriptstyle d<{r} <d+d' \atop d',\beta',\gamma'$}}  c_{21;{r},d'}^{\beta',\gamma',\beta} \e^{\rmi K
    \left(\frac{4{r} - d -2d'}{6} \right)}
  \ket{K,\beta'} \ket{{r}-d,\alpha}
  \ket{-{r}+d+d',\gamma'}   \\
  + \sum_{\makebox[0pt]{$\scriptstyle {r} >d+d' \atop d',\beta',\gamma'$}}  c_{21;{r},d'}^{\beta',\gamma',\beta}   \e^{\rmi K
    \left(\frac{4{r} - d -2d'}{6} \right)}
  \ket{K,\beta'} \ket{d',\gamma'} \ket{{r}-\left(d+d'\right),
    \alpha}    
\end{eqnarray}
\endnumpartsapp

\subsection{$H_{2}^{1}$}
The action of the operator $H_{2}^{1}$ on the two-triplon state $\ket{K,\alpha}\ket{d,\beta}$
is given by
\begin{eqnarray}
  H_{2}^{1} \ket{K,\alpha} \ket{d,\beta}  =& 
  \sum_{{r},\alpha'}  
  c_{12;{r},d}^{\alpha',\alpha,\beta}  \e^{\rmi K \left({r} + \frac{d}{2}
    \right) } \ket{K,\alpha'}
\end{eqnarray}
with
\begin{eqnarray}
  c_{12;r,d}^{\alpha',\alpha,\beta} :=&  \bra{R,\alpha'} H_{2}^{1}
  \ket{R+{r},\alpha} \ket{R+{r}+d, \beta} .
\end{eqnarray} 

The action of the operator $H_{2}^{1}$ on the three-triplon state $\ket{K,\alpha}\ket{d_1,\beta}\ket{d_2,\gamma}$
is given by
\numpartsapp
\begin{eqnarray}
  \fl H_{2}^{1} \ket{K,\alpha} \ket{d_1,\beta} \ket{d_2,\gamma}  
  = \sum_{\makebox[0pt]{$\scriptstyle {r} >-\left(d_1+d_2\right) \atop \alpha'$}} 
  c_{12;{r},d_1}^{\alpha',\alpha,\beta} \e^{\rmi K \left( \frac{3{r}+d_1-d_2}{6}\right)}
  \ket{K,\alpha'} \ket{{r}+d_1+d_2,\gamma}  \\
  + \sum_{\makebox[0pt]{$\scriptstyle {r} <-\left(d_1+d_2\right) \atop \alpha'$}} 
  c_{12;{r},d_1}^{\alpha',\alpha,\beta} \e^{\rmi K \left( \frac{3{r}+d_1-d_2}{6}\right)} \ket{K,\gamma} \ket{-\left({r}+d_1+d_2 \right),\alpha'}  \\
  + \sum_{\makebox[0pt]{$\scriptstyle {r} <d_1 \atop \beta'$}}
  c_{12;{r},d_2}^{\beta',\beta,\gamma} \e^{\rmi K \left(\frac{3{r}+ d_1 + 2
      d_2}{6} \right)}
  \ket{K,\alpha} \ket{-{r}+d_1,\beta'}  \\
  + \sum_{\makebox[0pt]{$\scriptstyle {r} >d_1 \atop \beta'$}}  c_{12;{r},d_2}^{\beta',\beta,\gamma} \e^{\rmi K \left(\frac{3{r}+ d_1 + 2
      d_2}{6} \right)} \ket{K,\beta'} \ket{{r}-d_1,\alpha}  \\
  +\sum_{\makebox[0pt]{$\scriptstyle {r} >-d_1 \atop \alpha'$}}
  c_{12;{r},d_1+d_2}^{\alpha',\alpha,\gamma} \e^{\rmi K \left(\frac{3{r}+ d_1 + 2
      d_2}{6} \right)}
  \ket{K,\alpha'} \ket{{r}+d_1,\beta}  \\
  +\sum_{\makebox[0pt]{$\scriptstyle {r} <-d_1 \atop \alpha'$}}
  c_{12;{r},d_1+d_2}^{\alpha',\alpha,\gamma} \e^{\rmi K \left(\frac{3{r}+ d_1 + 2
      d_2}{6} \right)} \ket{K,\beta} \ket{-\left({r}+d_1 \right),\alpha'} 
\end{eqnarray}
\endnumpartsapp

\subsection{$H_{1}^{3}$}
The action of the operator $H_{1}^{3}$ on the one-triplon state $\ket{K,\alpha}$
is given by
\begin{eqnarray}
     H_{1}^{3} \ket{K,\alpha}
     =& \sum_{\makebox[0pt]{$\scriptstyle {r},d_1',d_2' \atop \alpha',\beta',\gamma'$}} c_{31;{r},d_1',d_2'}^{\alpha',\beta',\gamma',\alpha} \e^{\rmi K\left({r}-\frac{2d_1'+d_2'}{3}\right)}
     \ket{K,\alpha'}\ket{d_1',\beta'}\ket{d_2',\gamma'}
\end{eqnarray}
with
\begin{eqnarray}
    c_{31;{r},d_1',d_2'}^{\alpha',\beta',\gamma',\alpha} := \bra{R,\alpha'}
    \bra{R+d_1',\beta'} \bra{R+d_1'+d_2',\gamma'} H_{1}^{3} \ket{{R+r},\alpha}.
\end{eqnarray} 

\subsection{$H_{3}^{1}$}
The action of the operator $H_{3}^{1}$ on the three-triplon state $\ket{K,\alpha}\ket{d_1,\beta}\ket{d_2,\gamma}$
is given by
\begin{eqnarray}
  H_{3}^{1} \ket{K,\alpha} \ket{d_1,\beta} \ket{d_2,\gamma} =& 
  \sum_{{r},\alpha'}  
  c_{13;{r},d_1,d_2}^{\alpha',\alpha,\beta,\gamma}  \e^{\rmi K \left({r} + \frac{2d_1+d_2}{3}
    \right) } \ket{K,\alpha'}
\end{eqnarray}
with
\begin{eqnarray}
  c_{13;r,d_1,d_2}^{\alpha',\alpha,\beta,\gamma}  
  :=  \bra{R,\alpha'} H_{3}^{1}
  \ket{R+{r},\alpha} \ket{R+{r}+d_1, \beta} \ket{R+r+d_1+d_2, \gamma}.
\end{eqnarray} 

\subsection{$H_{2}^{2}$}
The action of the operator $H_{2}^{2}$ on the two-triplon state $\ket{K,\alpha}\ket{d,\beta}$
is given by
\begin{eqnarray}
  H_{2}^{2} \ket{K,\alpha} \ket{d, \beta}
  =& \sum_{\makebox[0pt]{$\scriptstyle r,d' \atop \alpha',\beta'$}}  c_{22;{r},d',d}^{\alpha',\beta',\alpha,\beta}  \e^{\rmi K \left( {r} + \frac{d-d'}{2}\right)}\ket{K,\alpha'}
 \ket{d',\beta'}
\end{eqnarray}
with 
\begin{eqnarray}
  c_{22;{r},d',d}^{\alpha',\beta',\alpha,\beta} :=& \bra{R,\alpha'} \bra{R+d',\beta'} H_{2}^{2}
   \ket{R+r,\alpha} \ket{R+r+d,\beta} .
\end{eqnarray}

The action of the operator $H_{2}^{2}$ on the three-triplon state $\ket{K,\alpha}\ket{d_1,\beta}\ket{d_2,\gamma}$
is given by
\numpartsapp 
\begin{eqnarray}
  \fl H_{2}^{2} \ket{K,\alpha} \ket{d_1,\beta} \ket{d_2,\gamma}
  =  \sum_{\makebox[0pt]{$\scriptstyle {r} >- \left(d_1+d_2\right)+d_1' \atop d_1',\alpha',\beta'$}} c_{22;
    {r},d_1',d_1}^{\alpha',\beta',\alpha,\beta} \e^{\rmi K \left(\frac{2
      {r} + d_1 -d_1'}{3}\right)} 
  \ket{K,\alpha'} \ket{d_1',\beta'} \ket{{r} + d_1 +d_2 - d_1' ,
    \gamma} 
  \\
    {+}  \sum_{\makebox[0pt]{$\scriptstyle -\left(d_1+d_2\right)<{r} <- \left(d_1+d_2\right)+d_1' \atop d_1',\alpha',\beta'$}} c_{22;
      {r},d_1',d_1}^{\alpha',\beta',\alpha,\beta} \e^{\rmi K \left(\frac{2
        {r} + d_1 -d_1'}{3}\right)} \ket{K,\alpha'} \ket{{r} + d_1 +d_2,\gamma}
    \ket{-\left({r} + d_1 +d_2 \right)+d_1', \beta'}  
    \\
      {+}  \sum_{\makebox[0pt]{$\scriptstyle {r} <- \left(d_1+d_2\right) \atop d_1',\alpha',\beta'$}} c_{22;
        {r},d_1',d_1}^{\alpha',\beta',\alpha,\beta} \e^{\rmi K \left(\frac{2
          {r} + d_1 -d_1'}{3}\right)}  \ket{K,\gamma} \ket{-\left({r} +d_1 +d_2 \right),\alpha'}
      \ket{d_1',\beta'} 
      \\
      +  \sum_{\makebox[0pt]{$\scriptstyle -d_1<{r}<  d_2' \atop d_2',\alpha',\gamma'$}}  c_{22;
        {r},d_1+d_2',d_1+d_2}^{\alpha',\gamma',\alpha,\gamma} \e^{\rmi K \left(\frac{2{r} + d_2 -d_2'}{3}\right)}
      \ket{K,\alpha'} \ket{{r}+d_1,\beta} \ket{
        -{r}+d_2',\gamma'} 
      \\
      +  \sum_{\makebox[0pt]{$\scriptstyle {r}< - d_1 \atop d_2',\alpha',\gamma'$}}  c_{22;
        {r},d_1+d_2',d_1+d_2}^{\alpha',\gamma',\alpha,\gamma} \e^{\rmi K
        \left(\frac{2{r} + d_2 -d_2'}{3}\right)}
      \ket{K,\beta} \ket{-\left({r} + d_1 \right), \alpha'}
      \ket{d_1+d_2',\gamma'} 
      \\
      +  \sum_{\makebox[0pt]{$\scriptstyle {r}>  d_2' \atop d_2',\alpha',\gamma'$}}  c_{22;
        {r},d_1+d_2',d_1+d_2}^{\alpha',\gamma',\alpha,\gamma} \e^{\rmi K \left(\frac{2
          {r} + d_2 -d_2'}{3}\right)}
      \ket{K,\alpha'} \ket{d_1+d_2',\gamma'} \ket{{r}- d_2',\beta}  
      \\
      + \sum_{\makebox[0pt]{$\scriptstyle {r}< d_1 \atop d_2',\beta',\gamma'$}}  c_{22;
        {r},d_2',d_2}^{\beta',\gamma',\beta,\gamma}  \e^{\rmi K \left(\frac{2
          {r} + d_2 -d_2'}{3}\right)}
      \ket{K,\alpha} \ket{-{r}+d_1,\beta'} \ket{d_2',\gamma'} 
      \\
      + \sum_{\makebox[0pt]{$\scriptstyle d_1 < {r}< d_1+d_2' \atop d_2',\beta',\gamma'$}}  c_{22;
        {r},d_2',d_2}^{\beta',\gamma',\beta,\gamma}  \e^{\rmi K \left(\frac{2
          {r} + d_2 -d_2'}{3}\right)}  \ket{K,\beta'} \ket{{r}-d_1,\alpha} \ket{-{r}+d_1 +d_2' ,
        \gamma'} 
      \\
      + \sum_{\makebox[0pt]{$\scriptstyle {r}> d_1+d_2' \atop d_2',\beta',\gamma'$}}  c_{22;
        {r},d_2',d_2}^{\beta',\gamma',\beta,\gamma}  \e^{\rmi K \left(\frac{2
          {r} + d_2 -d_2'}{3}\right)} \ket{K,\beta'}\ket{d_2',\gamma'}
      \ket{{r} - \left(d_1 +d_2'\right),\alpha}
\end{eqnarray}
\endnumpartsapp

\subsection{$H_{2}^{3}$}
The action of the operator $H_{2}^{3}$ on the two-triplon state $\ket{K,\alpha}\ket{d,\beta}$
is given by
\begin{eqnarray}
  H_{2}^{3} \ket{K,\alpha} \ket{d,\beta} 
  = \sum_{\makebox[0pt]{$\scriptstyle {r},d_1',d_2' \atop \alpha',\beta',\gamma'$}} c_{32;{r},d_1',d_2',d}^{\alpha',\beta',\gamma',\alpha,\beta} \e^{\rmi K\left({r}+\frac{3d-4d_1'-2d_2'}{6}\right)}
  \ket{K,\alpha'}\ket{d_1',\beta'}\ket{d_2',\gamma'}
\end{eqnarray}
with
\begin{eqnarray}
\fl  c_{32;{r},d_1',d_2',d}^{\alpha',\beta',\gamma',\alpha,\beta}  := \bra{R,\alpha'}
    \bra{R+d_1',\beta'} \bra{R+d_1'+d_2',\gamma'} H_{2}^{3} \ket{{R+r},\alpha}
    \ket{R+r+d,\beta}.
\end{eqnarray} 

\subsection{$H_{3}^{2}$}
The action of the operator $H_{3}^{2}$ on the three-triplon state $\ket{K,\alpha}\ket{d_1,\beta}\ket{d_2,\gamma}$
is given by
\begin{eqnarray}
  H_{3}^{2} \ket{K,\alpha} \ket{d_1,\beta} \ket{d_2,\gamma} = 
  \sum_{\makebox[0pt]{$\scriptstyle {r},d' \atop \alpha',\beta'$}}  
  c_{23;{r},d',d_1,d_2}^{\alpha',\beta',\alpha,\beta,\gamma}  \e^{\rmi K \left({r} + \frac{4d_1+2d_2-3d'}{6}
    \right) } \ket{K,\alpha'} \ket{d',\beta'}
\end{eqnarray}
with
\begin{eqnarray}
\fl  c_{23;r,d',d_1,d_2}^{\alpha',\beta',\alpha,\beta,\gamma}
  :=  \bra{R,\alpha'} \bra{R+d',\beta'} H_{3}^{2}
    \ket{R+{r},\alpha} \ket{R+{r}+d_1, \beta} \ket{R+r+d_1+d_2, \gamma}.
\end{eqnarray}

\subsection{$H_{3}^{3}$}
The action of the operator $H_{3}^{3}$ on the three-triplon state $\ket{K,\alpha}\ket{d_1,\beta}\ket{d_2,\gamma}$
is given by
\begin{eqnarray}
\fl  H_{3}^{3} \ket{K,\alpha} \ket{d_1, \beta} \ket{d_2,\gamma}  
  = \sum_{\makebox[0pt]{$\scriptstyle r,d_1',d_2' \atop \alpha',\beta',\gamma'$}}
  c_{33;{r},d_1',d_2',d_1,d_2}^{\alpha',\beta',\gamma',\alpha,\beta,\gamma}
  \e^{\rmi K \left( {r} +
    \frac{2\left(d_1-d_1'\right)+\left(d_2-d_2'\right)}{3}\right)}\ket{K,\alpha'}
  \ket{d_1',\beta'} \ket{d_2',\gamma'}
\end{eqnarray}
with 
\begin{eqnarray}
  \eqalign{
  \fl  c_{33;{r},d_1',d_2',d_1,d_2}^{\alpha',\beta',\gamma',\alpha,\beta,\gamma}
  := \bra{R,\alpha'} \bra{R+d_1',\beta'} \bra{R+d_1'+d_2',\gamma'}H_{3}^{3}\\ \times \ket{R+r,\alpha} \ket{R+r+d_1,\beta} \ket{R+r+d_1+d_2,\gamma} .}
\end{eqnarray}

\subsection{$S=1$, $m=0$ subspace}
Due to the SU(2) it is possible to reduce the computational effort. The
one-, two- and three-triplon states with $S=1$ and $m=0$ are listed in
\tref{tab:subspace}. Since they are independent from the total momentum $K$
and the relative distances $d,d_1$ and $d_2$ we omit the dependence on these parameters.
\begin{table*}
  \caption{\label{tab:subspace} States of the $S=1$ and $m=0$ subspace.}
  \begin{indented}
    \item[]
    \begin{tabular}{@{}ll}
      \br
      $\ket{S=1,m=0}_{1\phantom{a}}$ & $ \ket{z} $ \\
      $\ket{S=1,m=0}_{2\phantom{a}}$ & $\frac{\rmi}{\sqrt{2}}\Big(\ket{x,y} -  \ket{y,x} \Big)$ \\
      $\ket{S=1,m=0}_{3a}$ & $-\sqrt{\frac{3}{20}}\Big(\ket{z,x,x} +
      \ket{z,y,y} + \ket{x,z,x}
      +  \ket{y,z,y}\Big)$\\
      &$-\sqrt{\frac{2}{30}}\Big(2\ket{z,z,z} -  \ket{x,x,z}
      -  \ket{y,y,z} \Big)$\\
      $\ket{S=1,m=0}_{3b}$ & $\frac{1}{2} \Big(   \ket{z,x,x}  +
      \ket{z,y,y} - \ket{x,z,x} -
      \ket{y,z,y}  \Big)$\\
      $\ket{S=1,m=0}_{3c}$ & $\frac{1}{\sqrt{3}} \Big( \ket{x,x,z} + \ket{y,y,z} +
      \ket{z,z,z} \Big)$ \\
      \br
    \end{tabular}
  \end{indented}
\end{table*}

\bibliographystyle{iopart-num.bst}

\section*{References}
\bibliography{bib}

\end{document}